\documentclass[%
 reprint,
 amsmath,amssymb,
 aps,
]{revtex4-1}

\usepackage{graphicx}
\usepackage{dcolumn}
\usepackage{bm}
\usepackage{xfrac}
\usepackage{siunitx}
\usepackage[hidelinks]{hyperref}

\usepackage{ulem}
\usepackage{upgreek}
\usepackage{siunitx}
\usepackage{xcolor}
\usepackage{caption}
\usepackage{subcaption}

\newcommand{\etal}{\textit{et al.}}

\newcommand{\black}{\color{black}}  

\begin{document}

\preprint{APS/123-QED}

\title{Exploring ultra-high-intensity wakefields in carbon nanotube arrays:\\ an effective plasma-density approach}

\author{A. Bonatto}
 \email{abonatto@ufcspa.edu.br}
 \affiliation{Graduate Program in Information Technology and Healthcare Management, and the Beam Physics Group, \\ Federal University of Health Sciences of Porto Alegre, Porto Alegre, RS, 90050-170, Brazil}

\author{G. Xia}
\author{O. Apsimon}
 \affiliation{%
 Department of Physics and Astronomy, The University of Manchester, Manchester, M13 9PL, United Kingdom\\
 and The Cockcroft Institute, Sci-Tech Daresbury, Warrington, WA4 4AD, United Kingdom
}%

\author{C. Bontoiu}
\author{E. Kukstas}
\author{V. Rodin}
\author{M. Yadav}
\author{C. P. Welsch}

\affiliation{
Department of Physics, The University of Liverpool, Liverpool, L69 3BX, United Kingdom\\
and The Cockcroft Institute, Sci-Tech Daresbury, Warrington, WA4 4AD, United Kingdom 
}%

\author{J. Resta-López}
\email{javier2.resta@uv.es}
\affiliation{ICMUV, Instituto de Ciencia de los Materiales, Universidad de Valencia, 46071 Valencia, Spain}

\date{\today}

\begin{abstract}
Charged particle acceleration using solid-state nanostructures has attracted attention in recent years as a method of achieving ultra-high-gradient acceleration in the TV/m domain. More concretely, metallic hollow nanostructures could be suitable for particle acceleration through the excitation of wakefields by a laser or a high-intensity charged particle beam in a high-density solid-state plasma. For instance, due to their special channelling properties as well as optoelectronic and thermo-mechanical properties, carbon nanotubes could be an excellent medium for this purpose. This article investigates the feasibility of generating ultra-high gradient acceleration using carbon nanotube arrays, modelled as solid-state plasmas in conventional particle-in-cell simulations performed in a two-dimensional axisymmetric (\textit{quasi}-3D) geometry. The generation of beam-driven plasma wakefields depending on different parameters of the solid structure is discussed in detail. Furthermore, by adopting an effective plasma-density approach, existing analytical expressions, originally derived for homogeneous plasmas, can be used to describe wakefields driven in periodic non-uniform plasmas. 
\end{abstract}

\maketitle


\section{Introduction}
\label{sec:intro}

High-energy particle accelerators are predominantly based on radiofrequency (RF) technology. However, standard RF technology is limited to gradients of the order of 100 MV/m due to surface breakdown \cite{breakdown}. Thus, larger and more expensive accelerator facilities are necessary in order to obtain higher energy particle beams. Therefore, R\&D into novel accelerator techniques is important to overcome the present acceleration limitations towards more compact and cost-effective solutions. Several alternative paths towards high-gradient acceleration are currently being investigated, e.g. techniques using dielectric microstructures \cite{DLA1, DLA2, DLA3} or plasmas as accelerating media. For instance, plasma wakefield acceleration (PWFA) methods based on gaseous plasma have been shown to produce gradients of up to approximately 100 GV/m \cite{PWFA1, PWFA2, PWFA3, PWFA4}. For typical gaseous plasmas used as acceleration media, the maximum achievable accelerating gradient is limited by the so-called plasma wave-breaking limit, which depends on the plasma density. In the linear regime, this limit, known as the cold non-relativistic wave breaking field \cite{Esarey:2009}, is given by 

\begin{align}
E_0[\text{V/m}]= m_e c \, \omega_p / e \simeq 96 \sqrt{n_0 [\text{cm}^{-3}]}, 
\label{eq:TDlimit}
\end{align}

\noindent where $m_e$ and $e$ are the electron mass and charge, respectively, $c$ is the speed of light in vacuum, $\omega_p = [n_0 e^2/(\epsilon_0 m_e)]^{1/2}$ is the plasma frequency, $n_0$ is the plasma density and $\epsilon_0$ is the vacuum permittivity. \black To surpass present PWFA limits, solid-based acceleration media, such as crystals or nanostructures could offer a solution. The density of charge carriers (conduction electrons) in solids is four or five orders of magnitude higher than those in a gaseous plasma, thus offering the possibility to
obtain ultra-high gradients on the order of $E_0 \sim 1\text{--}10$ TV/m, if the same linear theory is assumed.

Solid-state wakefield acceleration using crystals was proposed in the 1980s and 1990s by T. Tajima and others \cite{Tajima, Chen} as an alternative particle acceleration technique to sustain TV/m acceleration gradients. In the original Tajima's conceptual scheme, a metallic crystal is excited by a laser (laser driven), generating a longitudinal electric wakefield which can be used as an accelerating structure. Then, if a witness beam of charged particles is injected into the crystal with an optimal injection angle for channelling and with the right phase with respect to the wakefield, the channelled particles can experience acceleration. To reach accelerating gradients on the order of 1--10 TV/m, crystals must be excited by ultrashort X-ray laser pulses within a power range of TW-- PW, which makes the practical realisation of the concept very challenging. It has only recently become a realistic possibility since the invention of the single-cycled optical laser compression technique by G. Mourou et al. \cite{Mourou:2014}. Although simulation studies of X-ray wakefield acceleration have been performed \cite{Zhang}, it has not been experimentally demonstrated yet.
Alternatively, channelled particles in crystals can also be accelerated by means of electric wakefields excited by ultrashort, relativistic electron bunches (beam driven). In this case the energy losses of a driving bunch can be transformed into the acceleration energy of a witness bunch. 

If natural crystals (e.g. silicon) are used for the solid-state wakefield acceleration, the beam intensity acceptance is significantly limited by the angstrom-size channels. In addition, such small size channels increase the dechannelling rate and make channels physically vulnerable to high energy interactions, thus increasing the damage probability by high power beams. 

Over the past decade there have been great advances in nanofabrication techniques \cite{PingLi, Cary, XChen} that could offer an excellent way to overcome many of the limitations of natural crystals. Metallic nanostructures and metamaterials \cite{Pizzi, Ling-Bao} may offer suitable ultra-dense plasma media for wakefield acceleration or charged particle beam manipulation, i.e. channelling, bending, wiggling, etc. This also includes the possibility of investigating new paths towards ultra-compact X-ray sources \cite{Pizzi}.  

In this context, the use of nanotube structures for generating ultra-high gradients is attracting attention \cite{Shin1, Shin2, Resta}. For instance, carbon nanotube (CNT) based structures can help to relax the constraints to more realistic regimes with respect to natural crystals. CNTs present the following advantages with respect to natural crystals: (i) larger degree of dimensional flexibility and thermo-mechanical strength; (ii) transverse acceptances of the order of up to 100 nm, i.e. three orders of magnitude higher than a typical silicon channel; (iii) lower dechannelling rate; (iv) less disruptive effects such as filamentation and collisions. Therefore, CNTs are a robust candidate for solid-state wakefield acceleration. 

Wakefields in crystals or nanostructures can be induced by means of the excitation of high-frequency collective motion of conduction electrons through the crystalline ionic lattice. This collective oscillation of conduction electrons in metals, excited by external electromagnetic fields is commonly referred to as plasmon \cite{Sarid2010}. For instance, the excitation of surface plasmonic modes \cite{nanoplasmonics1, nanoplasmonics2, nanoplasmonics3, nanoplasmonics4}, driven either by charged particle beam \cite{Nejati, Stolckli, Wang, Mowbray} or by laser \cite{Zhang, Shin3}, could be used as a collective mechanism to generate high acceleration gradients in metallic nanotubes. To be effective, the driver dimensions should match the spatial ($\sim$nm) and time (sub-femtoseconds) scales of the excited plasmonic oscillations. Wakefield-driving sources working on these scales can be experimentally accessible nowadays or in the near future. For instance, attosecond X-ray lasers are possible thanks to the pulse compression technique invented by Donna Strickland and Gerard Mourou \cite{Mourou:1985}. In the case of beam-driven wakefields, future upgrades of the experimental facility FACET-II at SLAC \cite{Yakimenko1, Yakimenko2} might allow the access to "quasi-solid" and ultra-short electron beams, with densities up to $\sim 10^{24}$~cm$^{-3}$ and sub-micrometer bunch length scale. Recent studies have reported that ultra-short and high-density electron beams could lead to a nonlinear plasmonic regime, generating acceleration gradients beyond TV/m in micro- and nano-tubes. This is also known as the \textit{crunch-in} regime \cite{Sahai:2017, Sahai:2020, Sahai:2021} and could be a potential step towards the realisation of compact PeV colliders \cite{Sahai:2022}.

In this article we study the feasibility of generating ultra-high acceleration gradients in nanostructures based on CNTs. In addition, we show that, under proper conditions, by adopting an effective density, existing analytical estimates, originally derived for wakefields driven in homogeneous plasmas in the linear regime, can be used to describe the wakefields excited in such nanostructures. In particular, for multiple target configurations (single hollow plasma channel/CNT, CNT arrays) the amplitude of the longitudinal beam-driven wakefield is evaluated and compared to -- and shown to be in agreement with -- analytical estimates for the amplitude of longitudinal beam-driven wakefields, obtained by using an effective-density approach, as described along this work. Moreover, existing results \cite{Resta} for CNT beam-driven wakefields, previously simulated using 2D Cartesian geometry have been revisited with the code FBPIC (Fourier–Bessel Particle-In-Cell) \cite{Lehe:2016}, using a 2D axisymmetric geometry. Regarding the structure of this work, in Section~\ref{sec:2} the simulation model is described. Section~\ref{sec:3} investigates the case of beam-driven wakefields in single tubes, and the role of key parameters, such as tube wall thickness and aperture, is systematically studied. Section~\ref{sec:4} focuses on the case of an aligned multichannel structure, representing a CNT array. Highly-ordered nanotube bundles would allow to fabric macroscopic samples with transverse width on the order of centimetres, thus being able to cover all the transverse cross sections for beam waist-sizes on the order of tens and hundreds of micrometres. In this case, where we deal with an inhomogeneous plasma structure, the feasibility of using an effective-density approach is investigated. Finally, some conclusions are drawn in Section~\ref{sec:5}. 

\section{\label{sec:2} Simulation model}

Hollow plasma channels (HPC), consisting of cylindrical shells populated by a uniform, pre-ionized cold plasma of two species (ions and electrons), are adopted here as a first-order approximation to describe a CNT, or a larger structure made of CNT bundles, as shown in Fig. \ref{fig:simulascheme1}. The carbon ions are simulated as cold ions with mass $m_i = 12 \, m_p$, where $m_p$ is the proton mass, and charge $q_i = Z\,e = e$, where $Z$ is the atomic number and $e$ the fundamental charge. Because of the single-level ionization, for a given ion density $n_i$, the electron density $n_e$, initially cold as well, will have the same value ($n_e = n_i)$. The choice of $Z = 1$ was made aiming to obtain conservative, lower bound estimates for the wakefield amplitudes to be driven in carbon-based solid state plasmas.

Regarding the target geometry, two distinct configurations are investigated. First, ``large'' hollow plasma channels (HPC), with \si{\micro m}-wide apertures are used as targets. Such structures could be built with CNT bundles, as shown in Fig. \ref{fig:simulascheme1}, with much larger dimensions  than those of a nanostructured CNT, and thus capable of channeling $\sim$\si{\micro m} electron beams. In the second configuration, multiple concentric HPCs, with thicknesses (and gaps) of a few nm, are adopted to describe CNT array targets. For both cases, the walls are modelled as uniform plasmas, with an average, effective density, which is presented and discussed along this document.

Although this collisionless fluid model does not take into account the solid state properties emerging from the ionic lattice, such as, for example, the presence of polaritons, previous studies have shown that the wakefield formation and electron acceleration processes in crystalline structures are only slightly affected by the ionic lattice force \cite{Hakimi:2018}. Therefore, neglecting the ionic effects at a first approximation might be justified, and -- if this is the case -- conventional particle-in-cell (PIC) codes might be an useful tool to investigate ultra high-gradient acceleration, as well as plasmon modelling in solids \cite{Zhang,Sahai:2020,Ding:2020}. As it has been already shown~\cite{Ding:2020}, the PIC method can be very suitable to model solid-sate based plasmons, since it self-consistently solves the fields and the motion of a large assembly of charged particles for the required time ($\sim$ sub-fs) and spatial ($\sim$~nm) scales.

Due to the high computational cost of 3D PIC simulations, the 2D Cartesian geometry is often adopted. In such geometry, CNT walls are modelled as flat plasma sheets, with finite thickness and length, and infinite width. However, this geometry is known to affect the spatial derivatives of the fields \cite{Ngirmang:2016} if applied to describe a non-slab-like system. Given the close-to-cylindrical symmetry of the physical system under consideration, a PIC code with a spectral solver can provide an accurate 3D description of the system, at a computational cost similar to the cost of performing 2D Cartesian PIC simulations \cite{Lifschitz:2009}.

In this work, the Fourier-Bessel Particle-in-Cell (FBPIC) code \cite{Lehe:2016} is adopted to perform the simulations using the cylindrical CNT hollow plasma channel model. Although particles in FBPIC have 3D Cartesian coordinates, its solver uses a set of 2D radial grids, each of them representing an azimuthal mode $m$ ($m = 0,\, 1,\, \dots$). While the first mode ($m = 0$) describes axisymmetric fields, higher-order modes can be added to model departures from the cylindrical symmetry. For example, a linearly polarised laser can be computed by adding the mode $m = 1$. An interesting feature of the spectral solver implementation in FBPIC is the mitigation of spurious numerical dispersion, including the zero-order numerical Cherenkov effect \cite{Godfrey:1974}. \\

Compact high-energy electron beams are often reported in literature with dimensions ranging from a fraction to a few micrometers \cite{Yakimenko2,Albert:2020}. Therefore, in this work, beams with near-\si{\micro m} RMS sizes are used as drivers to excite the intense wakefields in hollow plasma channels, which are under investigation in this section. The beam driver is assumed to have a bi-Gaussian density profile,
\begin{align}
%
    n_b(\xi, r)/n_0 = \left( n_b/n_0 \right) \, e^{-\xi^2/(2\sigma_\xi^2)} \, e^{-r^2/(2\sigma_r^2)},
    \label{nb}
\end{align}
where $\xi \equiv z - ct$ is the beam co-moving coordinate, $c$ is the speed of light in vacuum, $n_b \equiv (Q/e) / [(2\pi)^{3/2} \sigma_\xi \sigma_r^2]$ is the peak beam-density, $n_0$ is the initial plasma electron density, $Q$ is the beam charge and $\sigma_\xi$, $\sigma_r$ are the beam longitudinal and radial RMS sizes, respectively. The beam has initial kinetic energy $E_{k0}$, and energy spread $\delta E_{k0}/E_{k0} = 1\%$. In addition,  $E_{k0}$ is chosen to ensure that the corresponding relativistic factor $\gamma$ satisfies the condition $\gamma \gg 1$, in order to increase the beam stiffness.

\section{\label{sec:3} Single tube in 2D axisymmetric geometry}

\begin{figure}[btb]
\includegraphics[width=8cm]{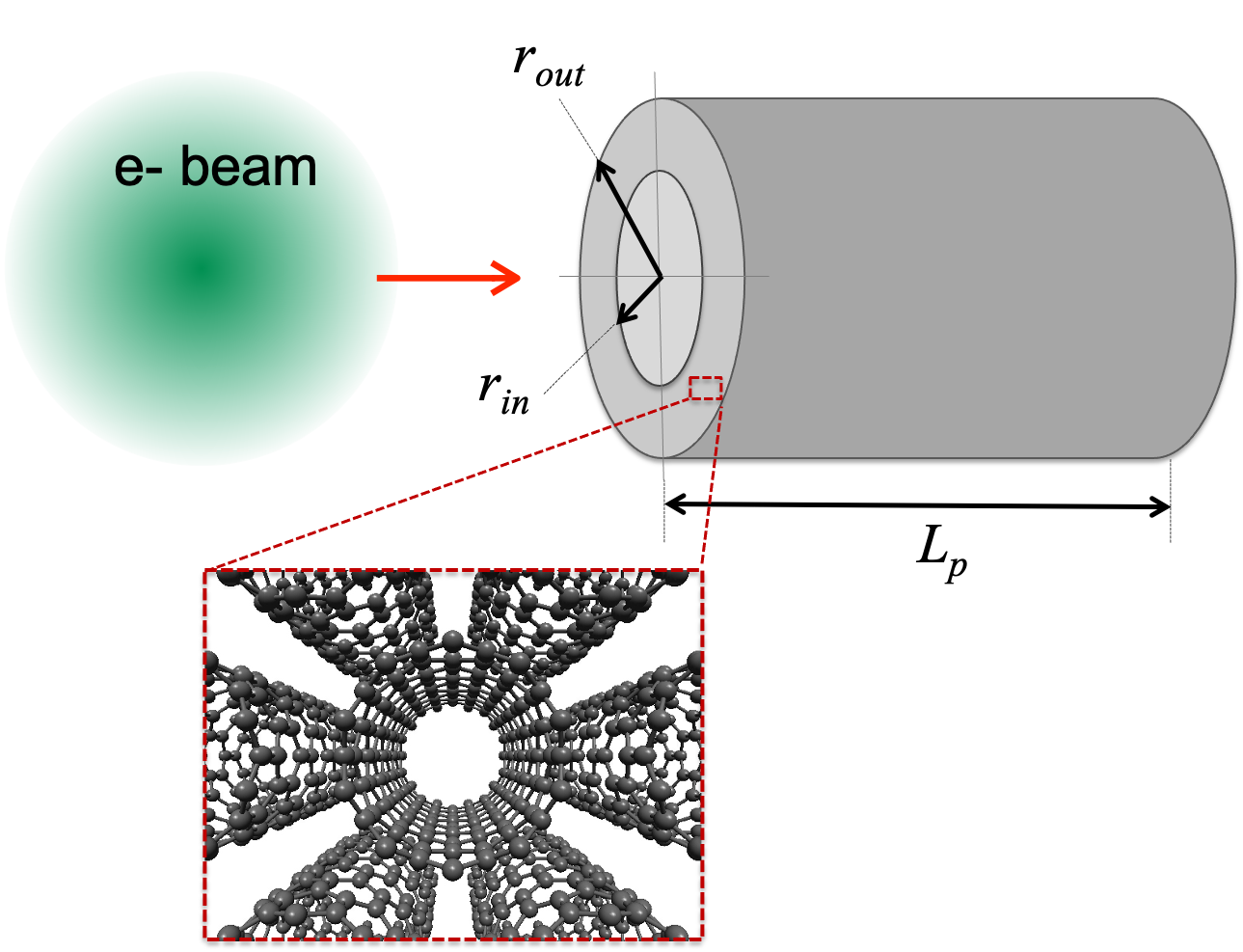}
\caption{Top: Schematic model for beam-driven wakefield simulation using a hollow cylinder of solid-state plasma confined in a wall of thickness $w = r_{\rm out}-r_{\rm in}$ and length $L_p$. Bottom: the cylinder wall could be made of CNT bundles (not to scale).}
\label{fig:simulascheme1}
\end{figure}

As a first approach, a single HPC is adopted as the medium for the beam-driven wakefield excitation. Figure~\ref{fig:simulascheme1} depicts a schematic of the system, in which an electron beam (driving source) is injected into a hollow plasma channel. The plasma is confined in the channel wall, assumed to be made up of CNT bundles. In principle, modern techniques allow for the fabrication of macroscopic materials based on aligned single and multi-wall CNT bundles or CNT forest films \cite{Zhu2004, Lalwani2012, Lange2021}. In this nanostructured materials, the density profile of the plasma can be controlled by the packaging configuration of the CNTs. By varying parameters from this structure, such as internal radius, wall thickness, and plasma density, it is possible to verify how the wakefield intensity is affected. In order to accommodate a near-\si{\micro m} beam, the hollow plasma channel also has a micrometer-scale. 

Typical electron densities ($n_e$) in solid-state plasmas lie within the range of $10^{19} \, \si{\centi m}^{-3} \le n_e \le 10^{24} \, \si{\centi m}^{-3}$ \cite{Chen:1987,Ostling:1997}. Aiming to maintain conservative estimates for the amplitude of the wakefields to be excited in these materials, the lower limit of this range is chosen as the initial density, $n_0 = 10^{19} \, \si{\centi m}^{-3}$, for both electrons and ions. In other words, $n_e = n_i = n_0$, where $n_i$ is the ion density. Although this density is much lower than that of a CNT wall ($\sim 10^{23} \, \si{\centi m}^{-3}$), it could represent electrons in gaps and hollow spaces of (partially ionized) targets made with CNT arrays or bundles. Moreover, for the chosen beam and plasma parameters, the wakefields are excited approximately in the linear regime. Hence, the obtained results can be scaled up to higher densities. If the plasma electron and peak beam densities ($n_e$ and $n_b$, respectively) are increased accordingly, then the ratio $E_z/E_0$, where $E_z$ is the longitudinal wakefield, should remain constant. In this case, the wakefield amplitude for a higher density can be estimated by multiplying the ratio $E_z/E_0$ obtained from the lower density simulation by the new value of $E_0$, calculated for the higher density.

For a density of $n_e = 10^{19} \, \si{\centi m}^{-3}$, a plasma wavelength of $\lambda_p = 10.6 \, \si{\micro m}$ is obtained. From now on, this quantity ($\lambda_p$) is adopted as the characteristic length scale to define the HPC and beam dimensions as follows. The HPC has a length $L_p = 10 \, \lambda_p$, internal radius $r_{in} = 0.1 \, \lambda_p$, external radius $r_{out} = 0.5 \, \lambda_p$, and wall thickness $w = r_{out} - r_{in} = 0.4 \, \lambda_p$. Regarding the beam, both the longitudinal and radial RMS sizes are $\sigma_\xi = \sigma_r = 0.1 \, \lambda_p$. For such dimensions, a charge of $Q = 33 \, \si{\pico C}$ is chosen, providing a normalized peak beam-density of $n_b/n_0 = 1.1$, i.e., right after the transition from an overdense to a underdense propagation in the plasma, in order to ensure that the beam will experience linear focusing forces \cite{Nakanishi:1991,Marocchino:2017}. The initial beam-energy is $E_{k0} = 1 \,\si{\giga eV}$, with an energy spread of $\delta E_{k0}/E_{k0} = 1\%$, and the transverse normalized beam emittance is null. Such parameters were chosen to produce a stiff beam, able to drive a stable wakefield along its propagation. Since the amplitude of the wakefield is evaluated and compared in multiple situations along the investigation, this is a relevant matter.

\begin{figure}[h]
\includegraphics[width=6.9cm]{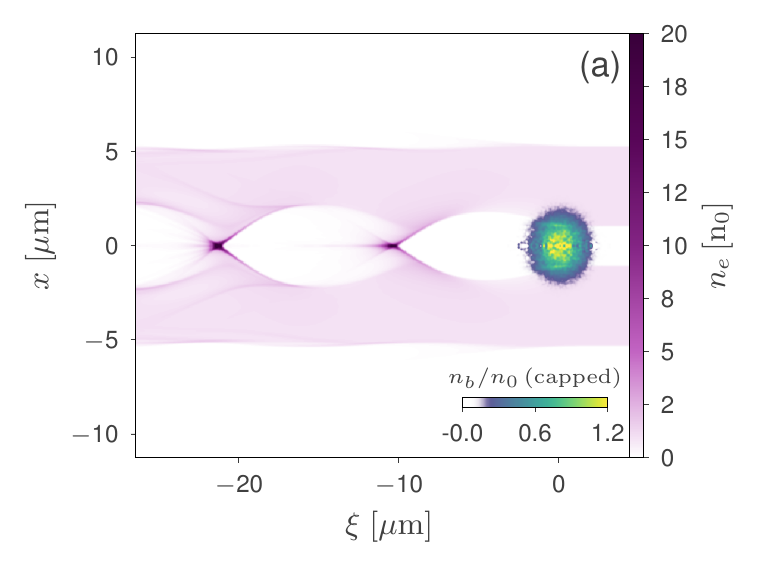} \\ \vspace{-0.4cm}
\includegraphics[width=7.4cm]{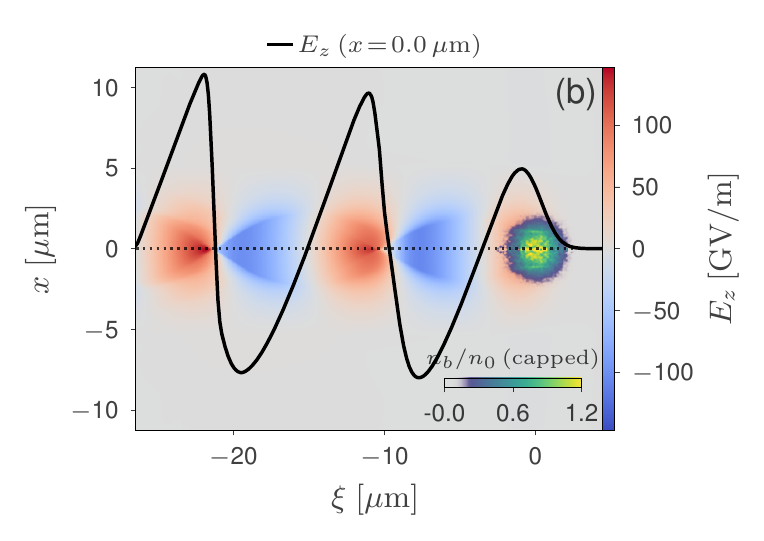} \\
\vspace{-0.4cm}
\includegraphics[width=7.4cm]{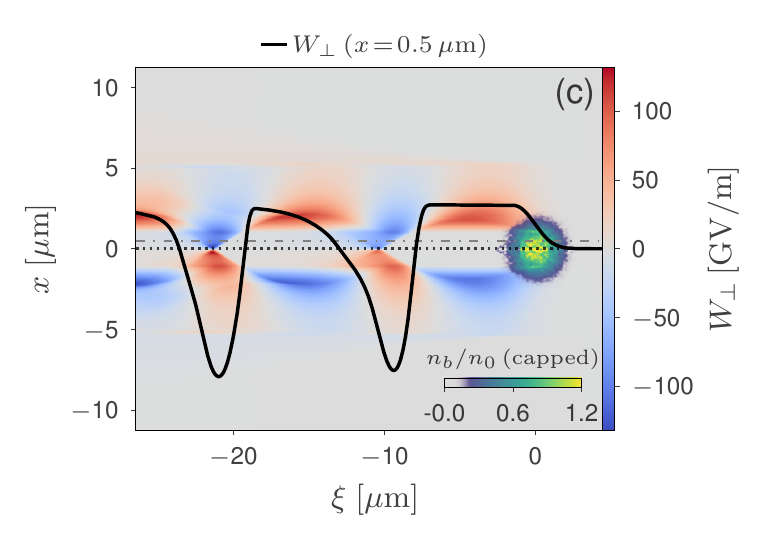} \vspace{-0.2cm}
\caption{Bi-Gaussian electron beam ($\sigma_\xi\! = \! \sigma_r \! = \! 0.1 \, \lambda_p$) propagating  along a hollow plasma channel with $r_{in} = 0.1 \,\lambda_p$, and $r_{out} = 0.5 \,\lambda_p$. Panel (a) shows the plasma-electron density (\textit{purple} color scale), and panels (b) and (c) depict the longitudinal and transverse beam-driven wakefields, $E_z$ and $W_\perp$, respectively, obtained from PIC simulation results (\textit{blue-grey-red} colored areas). The \textit{thick, solid black} lines represent $E_z$ on-axis ($x = 0$), and $W_\perp$ at $x = 0.5 \, \si{\micro m}$. The beam density (\textit{transparent-blue-green-yellow} color scale) is shown in all panels as well.}
\label{fig:single-CNT}
\end{figure}

Figure \ref{fig:single-CNT} shows PIC simulation results for the aforementioned parameters, taken at a propagation distance of $z = 53 \, \si{\micro m}$, corresponding to $L_p/2 = 5 \, \lambda_p$. Despite the betatron motion of individual beam particles, due to its high initial energy, the beam density profile remains mostly unchanged during the propagation. Figure \ref{fig:single-CNT}(a) depicts the beam (\textit{transparent-blue-green-yellow} color scale) and plasma-electron (\textit{purple} color scale) densities, in units of the initial plasma density $n_0$, both being saturated (capped) in order to enhance the visualization of how these densities are mutually affected. At this propagation distance, the beam density has been slightly modulated, being lower within the CNT walls due to the beam-wall interference. Regarding the plasma electrons, after being radially expelled by the beam, they experience a strong restoring force due to the carbon ions (not shown in this figure), which remain mostly undisturbed. As a consequence, these electrons are tightly focused, creating on-axis density spikes 20 to 35 times higher than the initial plasma density $n_0$, This behaviour has been described by Sahai \etal\, as the \textit{crunch-in} regime \cite{Sahai:2017, Sahai:2020, Sahai:2021}. The periodic transverse motion of plasma electrons, caused by the competition between the Coulomb repulsion and the restoring force due to the ions, creates an intense wakefield, which could be used as an accelerating structure for a witness beam. Figure \ref{fig:single-CNT}(b) shows the accelerating (negative) phase of the longitudinal wakefield $E_z(\xi, r)$ peaking at $|E_z^{max}| \simeq 105 \, \si{\giga V/m}$. This value is approximately one third of the cold non-relativistic wave breaking field $E_0$ calculated for the chosen density. The transverse wakefield, $W_\perp (\xi, r) = E_r - c B_\theta$, with $E_r$ the radial component of the electric field and $B_\theta$ the azimuthal component of the magnetic field, is shown in Fig.~\ref{fig:single-CNT}(c). While appreciable amplitudes can be seen within the CNT wall for both focusing and defocusing phases of the transverse wakefield, inside the CNT (i.e., for $r < r_{in}$) there are regions in which the transverse wakefield is approximately null. Due to this interesting feature, the use of hollow plasma channels to mitigate beam quality degradation caused by transverse effects is an active field of research \cite{Chiou:1995,Chiou:1998,Schroeder:1999,Kimura:2011,Adli:2016,Y.Li:2017}.

\subsection{\label{subsec:3.2.2} Tube aperture and wall thickness}

\begin{figure}[h!]
\includegraphics[width=7.5cm]{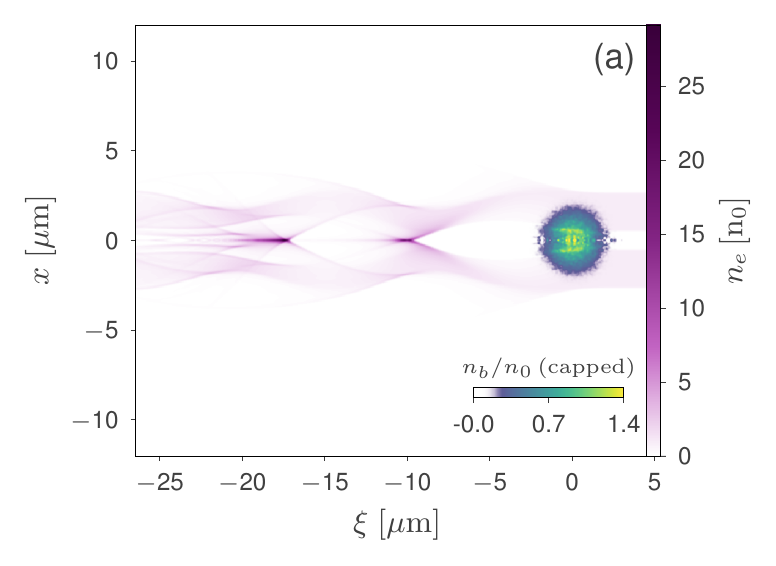}\\
\includegraphics[width=7.6cm]{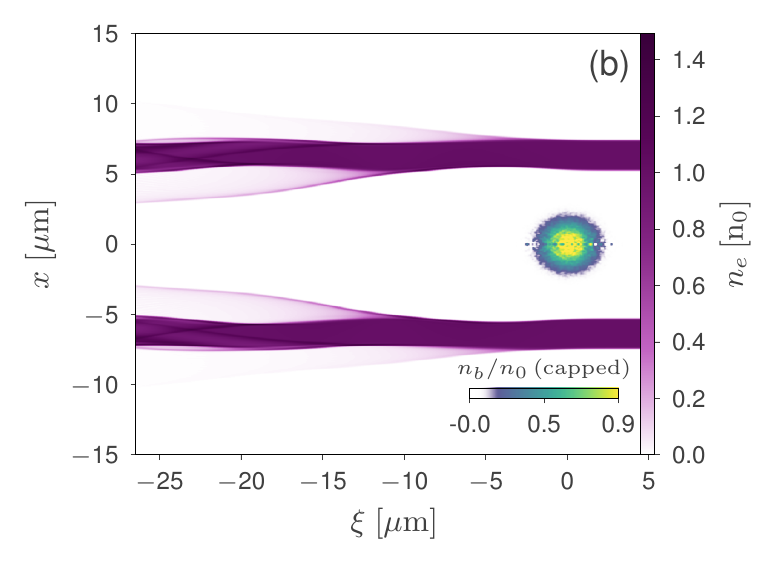}
\caption{Tube electron density (\textit{purple} color scale) and beam density (\textit{transparent-blue-green-yellow} color scale), capped at 75\% of its maximum value ($w = 0.2\, \lambda_p$, $z \simeq 70 \, \si{\micro m}$). (a) $r_{in} = 0.05 \, \lambda_p$, and (b) $r_{in} = 0.50 \, \lambda_p$.}
\label{fig:examplerin}
\end{figure}

Parameter scans might be helpful to determine the optimal system dimensions or aspect ratios to achieve as high amplitude wakefield as possible. In this sense, while maintaining the beam parameters and plasma density fixed, we have investigated the dependence of the longitudinal wakefield on both, tube aperture and wall thickness. In the first case, the inner tube radius $r_{in}$ has been varied. For instance, Fig.~\ref{fig:examplerin}, plotted for a fixed wall thickness $w = 0.2\, \lambda_p$, illustrates the tube electron density and beam density for the extreme (smallest and largest) investigated values of inner radius $r_{in}$, for a propagation distance of approximately $z = 70 \,  \si{\micro m}$. While in Fig.~\ref{fig:examplerin}(a), plotted for $r_{in} = 0.05 \, \lambda_p < \sigma_r \,(= 0.1 \, \lambda_p$), the beam transversely overlaps the tube wall, in Fig.~\ref{fig:examplerin}(b), plotted for $r_{in} = 0.50 \, \lambda_p > \sigma_r$, the tube inner radius is larger than the transverse beam size. The beam density color scale was saturated (capped) at 75\% of its maximum value, in order to improve the visualisation of how this quantity is affected by the interference with the tube wall. This saturation has been adopted for all beam density plots in this document. If the beam transversely overlaps the tube inner surface, (Fig. ~\ref{fig:examplerin}(a)) ionised electrons can move from the wall to the centre of the tube and form density bubbles, such as it is observed in a typical blowout regime or a nonlinear plasmonic regime \cite{Sahai:2017, Sahai:2020, Sahai:2021}. For a large tube aperture with respect to the transverse beam size (Fig.~\ref{fig:examplerin}(b)), with an initial normalized peak beam-density $n_b/n_0 \sim 1$, the excitation of electrons from the tube wall is much smaller and, in this case, we only expect the excitation of linear surface plasmonic modes.

\begin{figure}[h!]
\includegraphics[width=7.5cm]{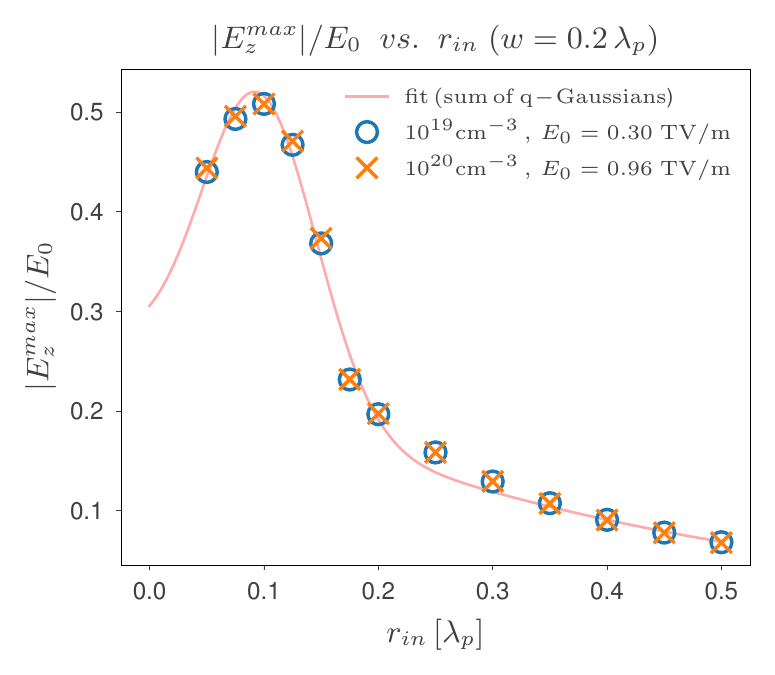}
\caption{Normalized maximum accelerating wakefield, $|E_z^{max}|/E_0$, plotted as a function of the internal radius $r_{in}$ in units of $\lambda_p$, for a  constant wall thickness $w = 0.2 \, \lambda_p$. Nearly identical PIC results were obtained for $n_e = 10^{19} \, \mathrm{cm}^{-3}$ (\textit{blue circles}) and $n_e = 10^{20} \, \mathrm{cm}^{-3}$ (\textit{orange crosses}). The obtained data can be fitted by a sum of two q-Gaussians (\textit{solid line}), written in Eq.~(\ref{eq:twoG}).}
\label{r_in}
\end{figure}

Results for the $r_{in}$ scan carried out with the FBPIC code are shown in Fig.~\ref{r_in}, in which the longitudinal wakefield amplitude $E^{max}_z$, taken from the accelerating wake phase -- and thus, negative -- and normalized by $E_0$, plotted as a function of the inner radius $r_{in}$. A maximum $|E^{max}_z|$ amplitude is obtained at $r_{in} \approx 0.1 \,\lambda_p$. For $r_{in} > 0.1 \,\lambda_p$ the longitudinal wakefield amplitude experiences a dramatic decrease, first a parabolic decline for $0.1\,\lambda_p < r_{in} < 0.16\, \lambda_p $, and a smoother decrease for $r_{in} > 0.16 \,\lambda_p$. Exactly the same behaviour is observed for two different plasma densities, $n_e=10^{19}$~cm$^{-3}$ and $n_e=10^{20}$~cm$^{-3}$. This is due to the fact that the results shown for each density in Fig. \ref{r_in} are normalized by their respective cold non-relativistic wave breaking fields, $E_0$. Such a scaling is expected to hold for wakefields driven in the linear to the quasi-linear regime \cite{Lu:2005}. It is interesting to observe that the PIC simulation points in Fig.~\ref{r_in} fit well to two q-Gaussian functions:

\begin{eqnarray}
 |E_z^{max}|/E_0 &\simeq& 0.32 \,e^{-197.8 (0.10\, -r_{in}/\lambda_p)^2} \nonumber \\
 && +7.80 \, e^{-0.42 (-r_{in}/\lambda_p-2.85)^2}.
 \label{eq:twoG}
\end{eqnarray}

\begin{figure}[h!]
\includegraphics[width=7.5cm]{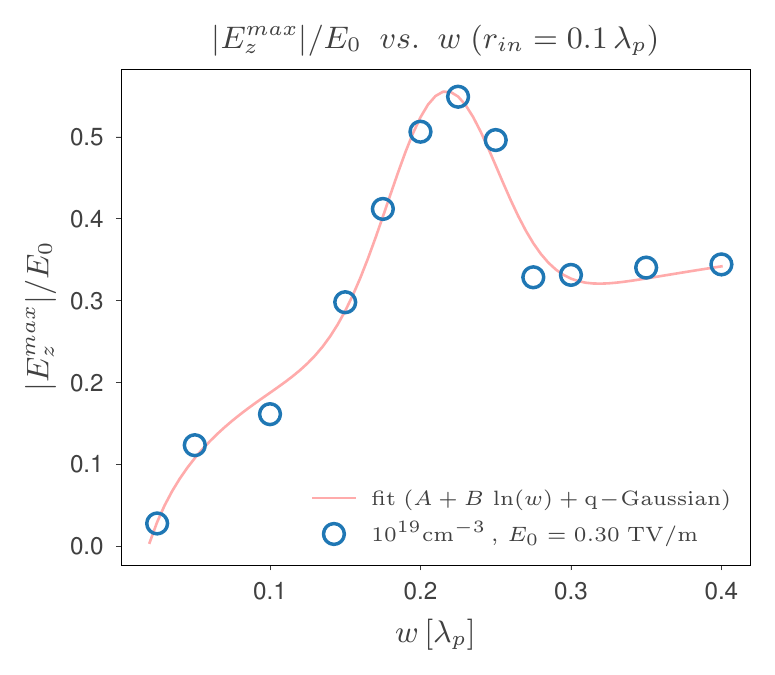}
\caption{Maximum accelerating wakefield $|E_z^{max}|/E_0$ vs. wall thickness $w = r_{out} - r_{in}$ obtained from PIC simulations for electron density $n_e = 10^{19} \, \mathrm{cm}^{-3}$ (\textit{blue circles}). The obtained simulation data can be well fitted by a sum of a logarithmic and a q-Gaussian function from Eq.~(\ref{eq:logandqG}) (\textit{solid line}).}
\label{wall}
\end{figure}

Another key parameter is the wall thickness $w=r_{out}-r_{in}$. Considering a plasma density of $n_e=10^{19}$~cm$^{-3}$ and a fixed internal channel radius set as $r_{in}=0.1 \, \lambda_p$, Fig.~\ref{wall} shows the result of the longitudinal amplitude as a function of $w$. A maximum gradient is obtained at $w \approx 0.22 \,\lambda_p$. The curve is very non-linear, and it seems to show an asymptotic behaviour for $w > 0.3 \,\lambda_p$, stabilising the values of the maximum achievable $|E^{max}_z|$ at a value of approximately $0.35 \, E_0$, with $E_0=0.3$~TV/m. In this case the simulation results show a good fit to the following function:
\begin{eqnarray}
 |E_z^{max}|/E_0 & \simeq & 0.28 \, e^{-384.67(0.21 - w/\lambda_p)^2} \nonumber \\ 
 && + 0.11 \ln(w/\lambda_p)+ 0.45 . 
 \label{eq:logandqG}
\end{eqnarray}
Eventually, expressions such as (\ref{eq:twoG}) and (\ref{eq:logandqG}) might allow one to make relatively quick estimates of the longitudinal wakefield amplitude as a function of inner radius and wall thickness, respectively, for these kinds of tubular structures.

\section{\label{sec:4} CNT array in 2D axisymmetric geometry}

Micrometer-scale targets, built with arrays of many coupled 2D carbon-based materials, e.g. highly-ordered CNT arrays or porous alumina, could be used as targets for channeling \si{\micro m}-scale electron beams, allowing for the whole transverse beam cross section to be covered. In the following subsection, an effective density is defined for periodically inhomogeneous structures, made of alternating uniform plasma-density layers (representing, for example, CNT walls) and vacuum gaps (representing the spaces between the material layers). Moreover, an investigation is presented on how to use this effective density to obtain estimates for the wakefields driven in such inhomogeneous structures.

\subsection{\label{subsubsec:3.2.1} Effective density}

\begin{figure}[htb]
\includegraphics[width=7cm]{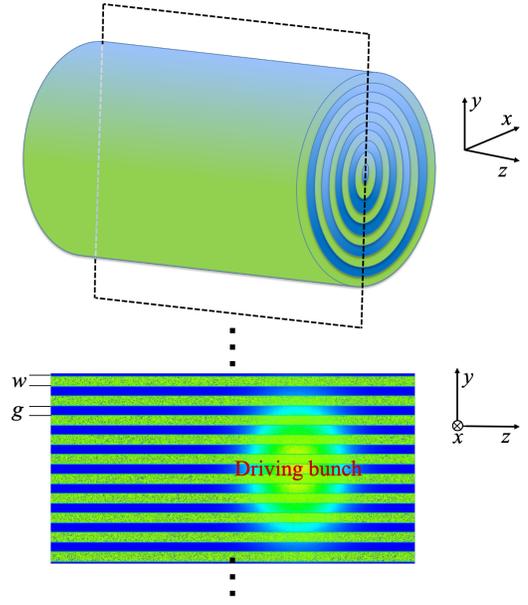} 
\caption{Schematic of the 2D axisymmetric model of the CNT array.}
\label{fig:CNTarraymodel}
\end{figure}

Analytical results from the linear theory can be used to describe small-amplitude, laser or beam-driven wakefields in uniform plasmas \cite{Esarey:2009}. Excitation of such wakefields are due to macroscopic, collective effects, caused by the displacement of plasma electrons by the wakefield driver. Hence, under proper conditions, the effect of homogeneous, spatially periodic microscopic departures from uniformity in plasma distribution may be smoothed out in the obtained wakefield. Under such conditions, it might be possible to define an effective density $n_\mathit{eff}$, to be used in the existing analytical expressions, derived for uniform plasmas, to describe wakefields driven in plasmas with small, periodic departures from uniformity. If this is the case, then the effective density can be interpreted as the average density of a unitary plasma cell, i.e., the smallest plasma volume (or surface, if a 2D Cartesian geometry is assumed)  which contains a single occurrence of the aforementioned periodic plasma-density pattern. Taking advantage of the 2D axisymmetric geometry adopted for the simulation, the validity of this hypothesis is here verified for a periodic CNT array, composed by multiple concentric cylinders, each of them having the same wall thickness $w$, separated by gaps with constant width $g$, as shown in Fig.~\ref{fig:CNTarraymodel}. Such arrangement creates alternating spatial regions, containing either a constant plasma-density $n_0$, or vacuum. The longitudinal section of such structure can be seen as a periodic distribution of constant-density plasma layers and gaps. By taking a single unit of this periodic pattern, i.e., one plasma layer, with constant density $n_0$ and thickness $w$, and one gap, with ``thickness'' (width) $g$, the effective density $n_\mathit{eff}$ can be defined as follows:
\begin{align}
    n_\mathit{eff} = n_0 \,/\, \kappa \;\;,\;\; \kappa \equiv (w + g)/w \,.
    \label{eq:n_eff}
\end{align}
Since the wall and gap thicknesses ($w$ and $g$, respectively) are non-negative quantities, from Eq.~(\ref{eq:n_eff}) one can see that the upper bound for the effective density is $n_\mathit{eff} = n_0$, which is reached for a uniform plasma ($g = 0$, $\kappa = 1$). As gaps are added, the larger the gap with respect to the wall thickness, the lower the effective density. The reciprocal of the coefficient defined in Eq.~(\ref{eq:n_eff}), i.e., $1/\kappa$, can be related to the fraction of the target volume that is filled with constant-density plasma, $V_\mathit{fill}$. For a planar geometry (for example, an array of graphene layers), the relation is straightforward, and $V_\mathit{fill} = 1/\kappa$. For a cylindrical geometry (for example, an array of concentric CNTs), the fraction of the filled volume is given by
\begin{align}
    V_\mathit{fill}(N,\kappa) = \left[ \frac{(N+1)}{N} - \frac{1}{N\kappa} \right]\frac{1}{\kappa} \,,
    \label{eq:Vfill}
\end{align}
where $N \equiv r_\mathit{out}/(w+g)$ is the number of ``repetitions'' of the quantity $(w+g)$ within the CNT array maximum radius, $r_\mathit{out}$. It is worth noting that, for a large number of layers and gaps, $V_\mathit{fill} \simeq 1/\kappa$, i.e., Eq.~(\ref{eq:Vfill}) converges to the same expression that describes the fraction of filled volume for a planar geometry. For example, for a large number of evenly distributed plasma layers and gaps ($w = g$), $\kappa = 2$, and $n_\mathit{eff} = n_0/2$. Since $1/\kappa = 0.5$, half of the target volume is occupied by plasma with constant-density $n_0$, and the other half is empty.

Existing analytical estimates for wakefields in homogeneous plasmas can be adapted to describe such fields in CNT arrays. This can be achieved by replacing the homogeneous plasma density $n_0$ in such estimates by the effective density $n_\mathit{eff}$. For example, considering a bi-Gaussian electron beam as the driving source, with density profile given by Eq.~(\ref{nb}), from the linear perturbation theory of PWFA, the amplitude of the longitudinal wakefield can be calculated from the following expression \cite{Esarey:2009}:

\begin{eqnarray}
 E^{max}_{z}&=&\sqrt{\frac{\pi}{2}}E_{0,\mathit{eff}} \, k^3_{p, \mathit{eff}} \, \sigma_{\xi} \sigma^2_r \left(\frac{n_b}{n_\mathit{eff}}\right)e^{-k^2_{p, \mathit{eff}}\frac{(\sigma^2_{\xi}-\sigma^2_r)}{2}} \nonumber \\
 && \times \Gamma\left(0, k^2_{p, \mathit{eff}} \, \sigma^2_r/2\right),
 \label{eq:amplitude}
\end{eqnarray}

\noindent where $k_{p, \mathit{eff}} = \omega_{p,\mathit{eff}} /c$ is the plasma wavenumber, calculated using $n_\mathit{eff}$, $E_{0,\text{{eff}}} = m_e c \, \omega_{p,\mathit{eff}} / e$ is the cold non-relativistic wave breaking field given associated to $n_{\mathit{eff}}$, and $\Gamma(0, k^2_{p, \mathit{eff}} \, \sigma^2_r/2)$ the incomplete gamma function. 

\begin{figure}
\includegraphics[width=7.5cm]{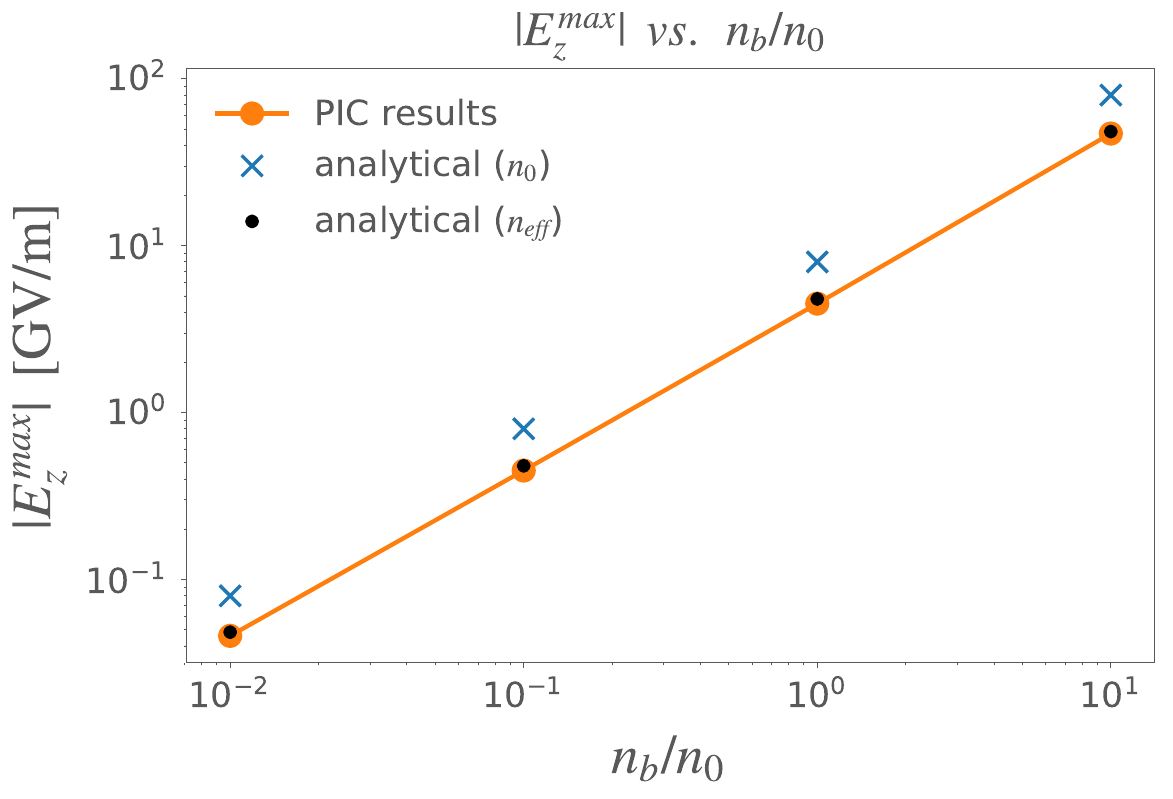}
\caption{Numerical and analytical estimates for the amplitude of the longitudinal wakefield $E_z^\text{max}$, for distinct values of $n_b/n_0$. While estimates obtained assuming $n_0$ as the plasma density (\textit{blue crosses}) are overestimated, estimates based on $n_\mathit{eff}$ (\textit{black dots}) show a remarkable agreement with PIC simulation results (\textit{orange line}).}
\label{fig:n_eff}
\end{figure}

Figure~\ref{fig:n_eff} compares the amplitude of the first accelerating phase of the longitudinal wakefield, $E_z^\text{max}$, obtained by means of PIC simulations and analytical estimates from Eq. (\ref{eq:amplitude}). The wakefield is driven by a bi-Gaussian electron beam propagating in a CNT array. This array has a length of $100 \, \si{\micro m}$, and it is composed by 25 concentric cylindrical layers, each of them having homogeneous density $n_0 = 10^{19} \, \si{\centi m^{-3}}$ and thickness $w = 40 \, \si{\nano m}$, separated by gaps of equal size (i.e., $g = 40 \, \si{\nano m}$). The first CNT layer has an internal radius of $20 \, \si{\nano m}$. The beam has longitudinal and transverse RMS sizes of $\sigma_\xi = 0.5/k_p \approx 0.84 \, \si{\micro m}$ and $\sigma_r = 0.1/k_p \approx 0.17 \, \si{\micro m}$, respectively. The initial beam energy is $E_{k0} = 1 \, \si{\giga eV}$ with an energy spread of $\delta E_{k0}/E{k0} = 1\%$. The beam charge of each simulation was chosen to provide one of the following values for the normalized peak beam-density, $n_b/n_0 = \{ 0.01\,,\, 0.1 \,,\, 1 \,,\, 10\}$. For these values, the beam-driven wakefields will be excited from the linear to the quasi-linear regime. Indeed, despite being derived for small values of $n_b/n_0$, the analytical expressions that describe beam-driven wakefields in the linear regime are known to hold reasonably well for $n_b/n_0 \lesssim 10$ \cite{Lu:2005}. Simulations were performed with a spatial domain of $17 \, \si{\micro m} \times 2.1 \si{\micro m}$ ($\xi \times r$), with longitudinal and transverse resolutions of $\sigma_z/25 \simeq 0.83 \, \si{\micro m}$ and $\sigma_r/25 \simeq 7 \, \si{nm}$, respectively. The only exceptions are the simulations in which the gaps have 5 nm; in these cases, a transverse resolution of $1 \, \si{nm}$ was adopted. Due to the cylindrical symmetry, a single azimuthal mode ($m=0)$ was used. Moreover, a total density of $36$ particles per cell was adopted. It can be seen in Fig.~\ref{fig:n_eff} that, while the analytical estimates for $E_z^\text{max}$ obtained assuming $n_0$ as the plasma density \cite{Esarey:2009} are overestimated, estimates obtained by using $n_\mathit{eff}$ show a remarkable agreement with simulation results, all along the investigated range.

\begin{figure}
\includegraphics{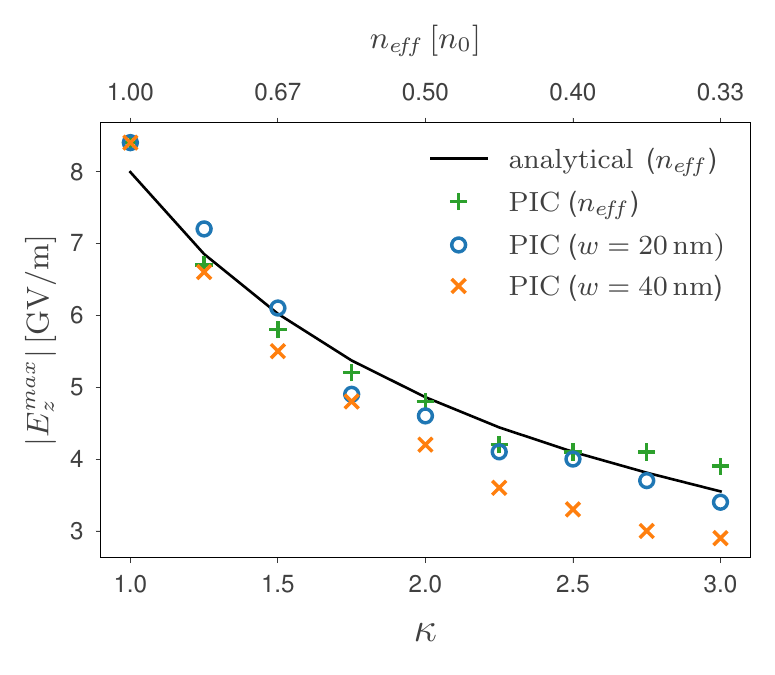}
\caption{Numerical and analytical estimates of $E_z^\text{max}$ for distinct values of $\kappa$. Analytical (\textit{black line}) and PIC (\textit{green crosses}) results obtained for a uniform plasma with effective density $n_\mathit{eff}$ are in agreement with PIC results for a CNT array with a wall thickness of $w = 20 \, \si{\nano m}$ (\textit{blue circles}). For $w = 40 \, \si{\nano m}$ (\textit{orange cross markers}), the agreement decreases as $\kappa$ is increased.}
\label{fig:n_eff_2}
\end{figure}

In order to further explore the robustness of analytical estimates based on the effective density, new comparisons are conducted against additional sets of PIC simulations, performed for distinct configurations of CNT arrays. For two distinct wall thicknesses, $w = 20 \, \si{\nano m}$ and $40 \, \si{\nano m}$, respectively, the gap width $g$ is varied in order to obtain values of $\kappa$ ranging from 1 ($n_\mathit{eff} = n_0$) to 3 ($n_\mathit{eff} = n_0/3$). A beam charge of $Q \approx 0.6 \, \si{\pico C}$ is set to provide a normalized peak beam-density of $n_b/n_0 = 1$. All the remaining parameters are the same as previously described. Results of these additional simulations, taken at a propagation distance of $z = 60.5 \, \si{\micro m}$, can be seen in Fig. \ref{fig:n_eff_2}. As shown in this figure, the analytical estimates of $E_z^\text{max}$, obtained for a homogeneous plasma with an effective density $n_\mathit{eff}$, are in agreement with PIC simulation results for a CNT array with a wall thickness of $w = 20 \, \si{\nano m}$. Despite minor departures in the intermediary region of the investigated range of $\kappa$ values, this agreement holds quite well, even for an effective density as low as $n_\mathit{eff} = n_0/3$. On the other hand, for a CNT array with $w = 40 \, \si{\nano m}$, PIC simulation results show a departure from the analytical estimates of $E_z^\text{max}$ that increases with $\kappa$. This behaviour might be explained as follows. Since the coefficient $\kappa$ is proportional to the ratio between wall and gap thicknesses, for the same value of $\kappa$, a CNT array with thicker walls will have wider gaps as well. While the gaps are thin enough to be ``populated'' by electrons from its adjacent walls, and while such gaps are small if compared to the beam transverse size, the effective-density model is likely to work fine. However, wide gaps between consecutive CNT-array layers may lead to plasma discontinuities that cannot be smoothed out by the plasma collective effects. Once the interaction between consecutive walls fade, or once the gaps are wide if compared to the beam transverse size, the idea of using an effective density to obtain analytical estimates for the beam-driven wakefield in the CNT array may no longer holds. \black 

\begin{figure*}
    \centering
    \includegraphics[scale=1.]{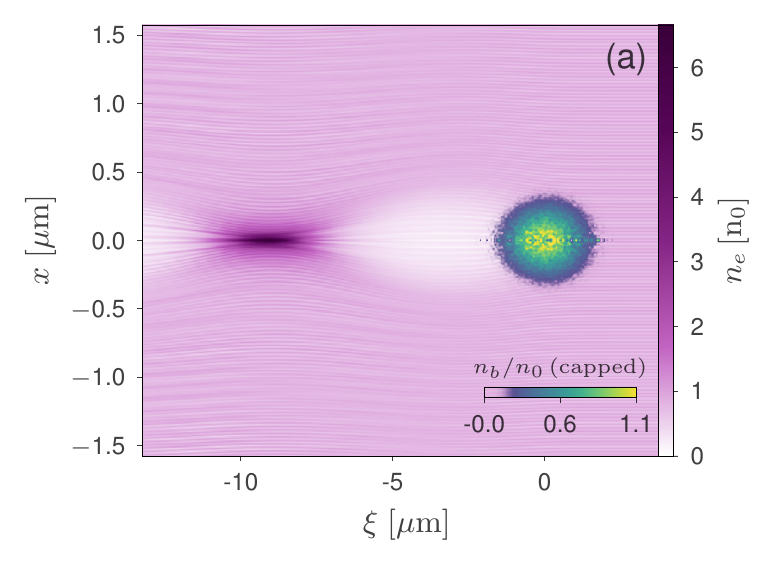}
    \includegraphics[scale=1.]{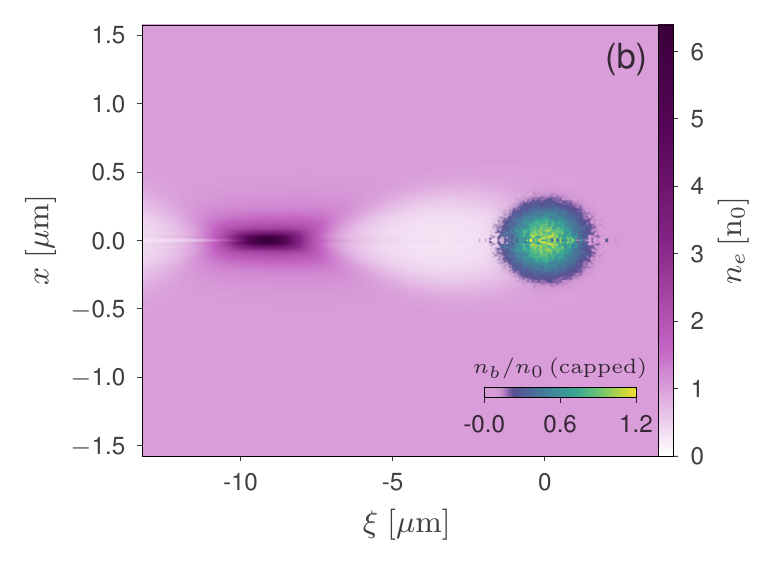} \\
    \includegraphics[scale=1.]{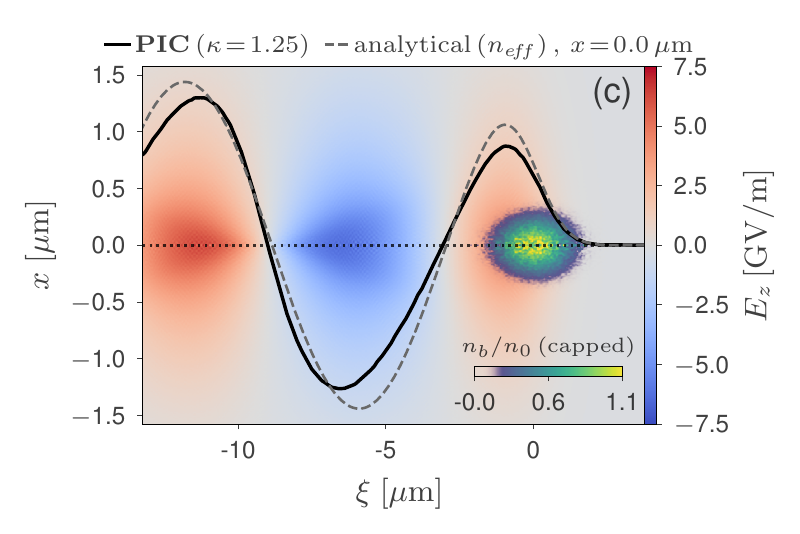}
    \includegraphics[scale=1.]{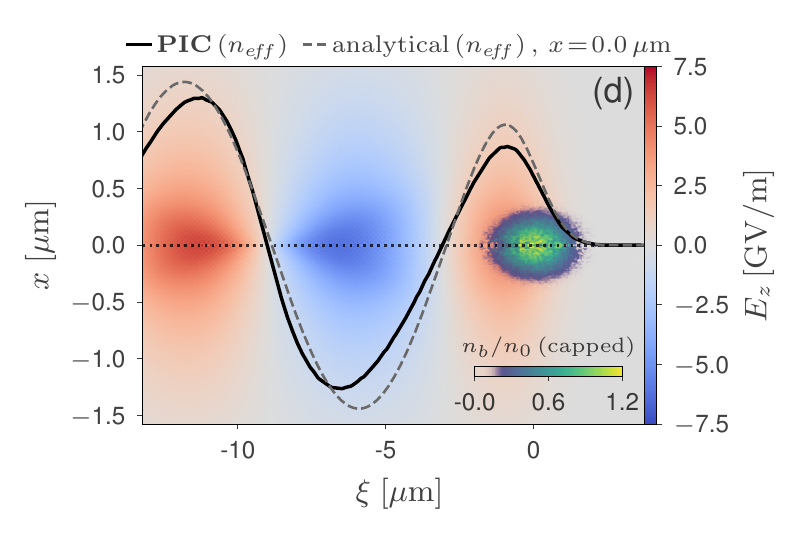} \\
    \includegraphics[scale=1.03]{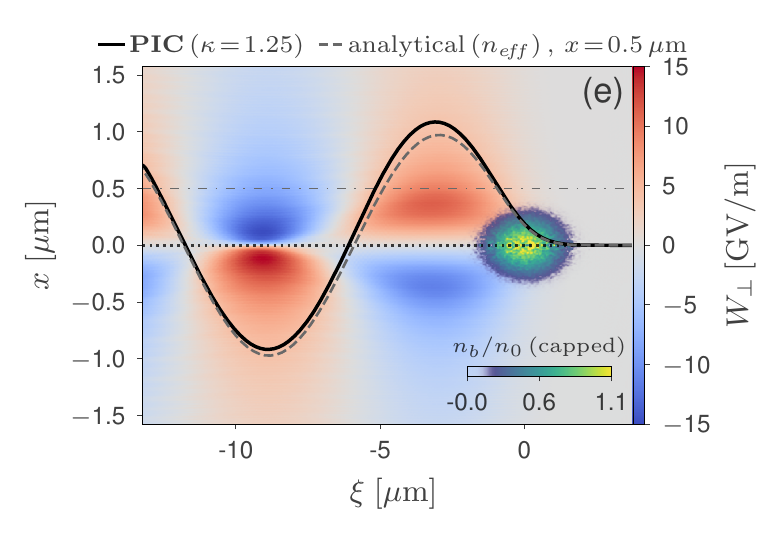}
    \includegraphics[scale=1.03]{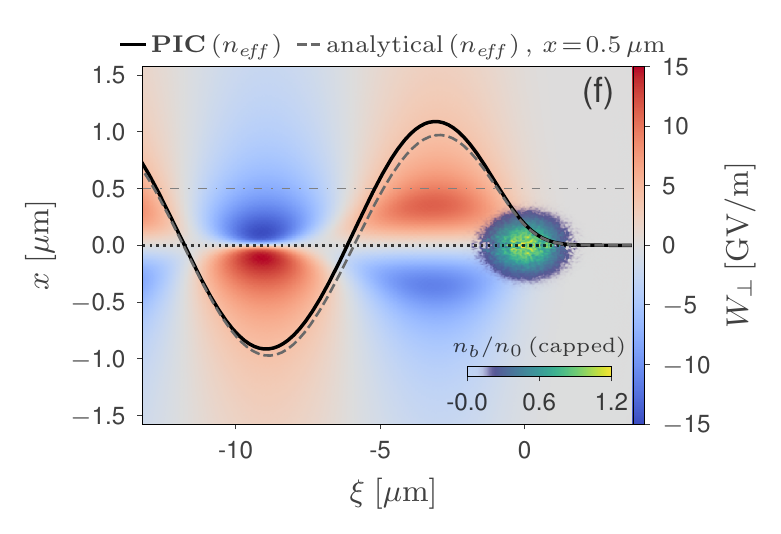}
    \caption{Simulation results for a CNT array with $w = 20 \, \si{nm}$ and $g = 5 \, \si{nm}$ ($\kappa = 1.25$, \textit{left column}), compared to  simulation results for a homogeneous plasma, with $n_\mathit{eff} = 0.8\,n_0$ (\textit{right column}), at a propagation distance of $z = 60 \, \si{\micro m}$. Panels (a) and (b) show the normalized plasma electron density (\textit{purple} color scale), panels (c) and (d) depict the longitudinal wakefields, and panels (e) and (f) illustrate the transverse wakefields. The fields obtained from PIC simulation results are shown as \textit{blue-grey-red} colored regions, with \textit{thick, solid black} lines representing the on-axis ($x = 0$) outline of the longitudinal wakefields, and the off-axis ($x = 0.5 \, \si{\micro m}$) outline of the transverse wakefields. Outlines for these same fields, obtained from effective-density-based analytical estimates (\textit{light, dashed black} lines), are provided as well. All panels show the beam density distribution (\textit{transparent-blue-green-yellow} color scale), capped (saturated) to emphasize internal perturbations caused by the CNT array layers.}
    \label{fig:w20_g05}
\end{figure*}

\begin{figure*}
    \centering
    \includegraphics[scale=1.]{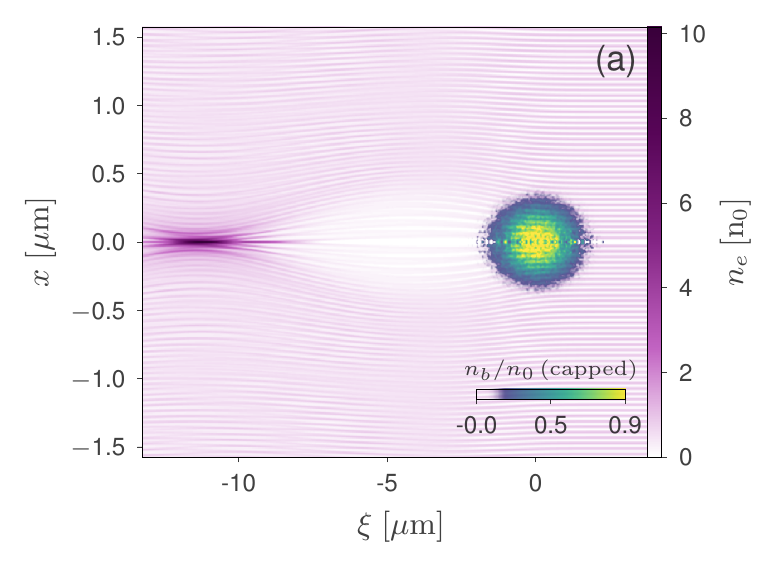}
    \includegraphics[scale=1.]{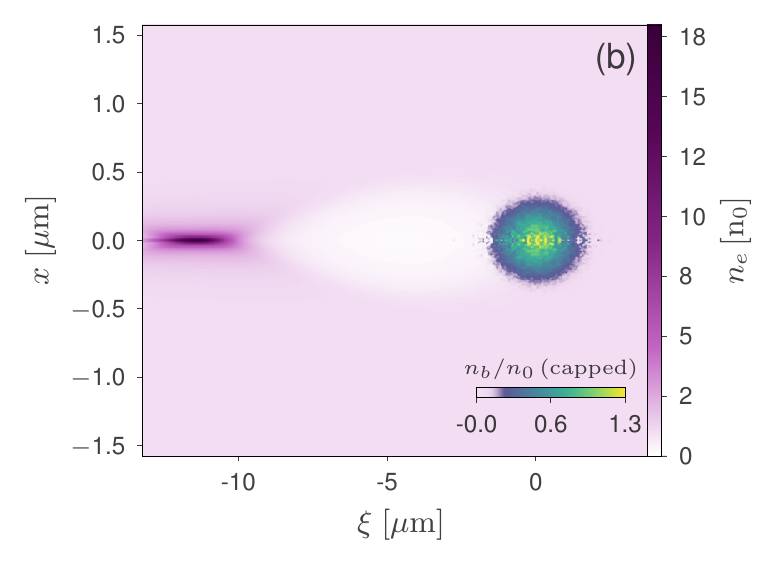} \\
    \includegraphics[scale=1.]{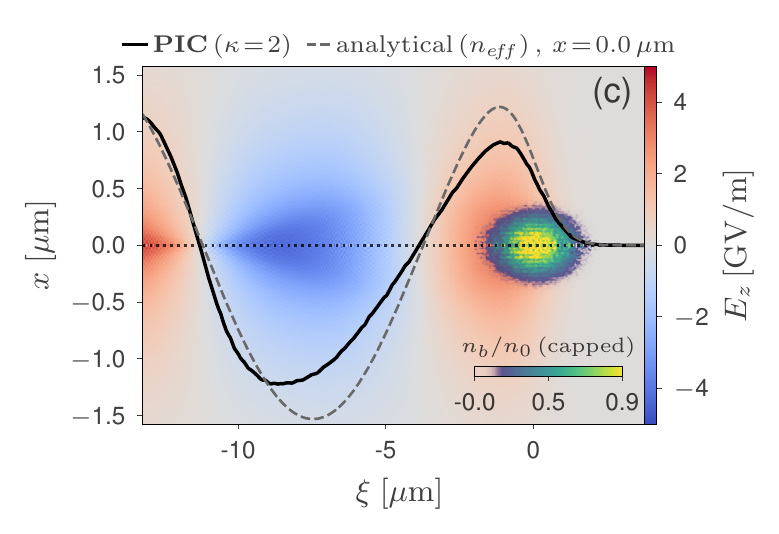}
    \includegraphics[scale=1.]{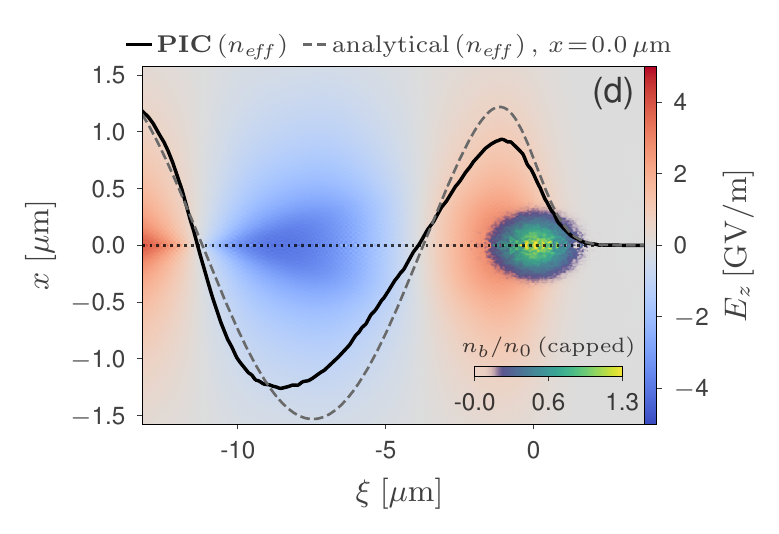} \\
    \includegraphics[scale=1.03]{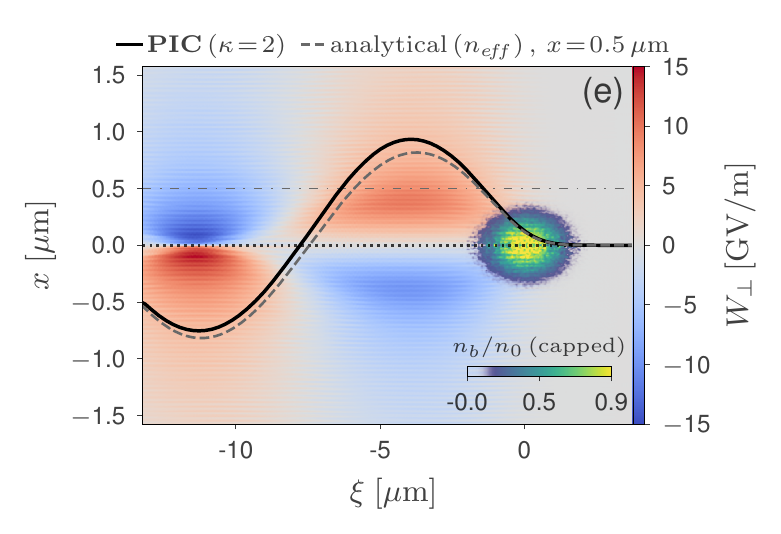}
    \includegraphics[scale=1.03]{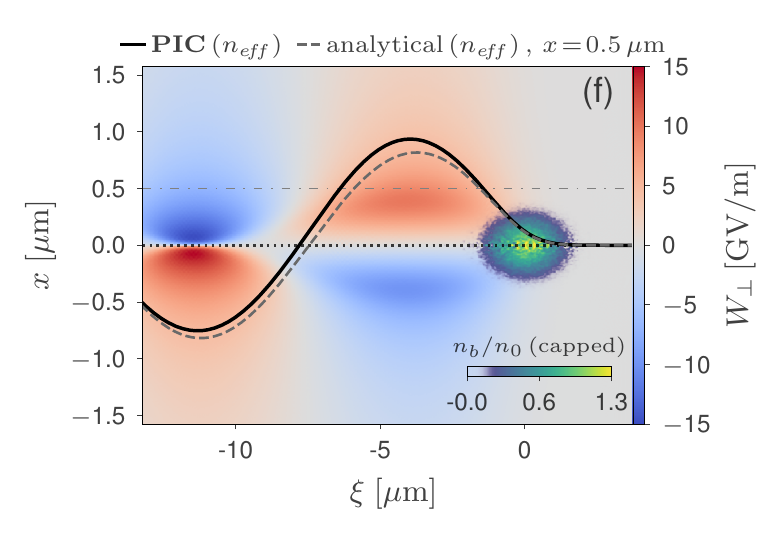}
    \caption{Simulation results for a CNT array with $w = 20 \, \si{nm}$ and $g = 20 \, \si{nm}$ ($\kappa = 2$, \textit{left column}), compared to  simulation results for a homogeneous plasma, with $n_\mathit{eff} = 0.5\,n_0$ (\textit{right column}), at a propagation distance of $z = 60 \, \si{\micro m}$. Panels (a) and (b) show the normalized plasma electron density (\textit{purple} color scale), panels (c) and (d) depict the longitudinal wakefields, and panels (e) and (f) illustrate the transverse wakefields. The fields obtained from PIC simulation results are shown as \textit{blue-grey-red} colored regions, with \textit{thick, solid black} lines representing the on-axis ($x = 0$) outline of the longitudinal wakefields, and the off-axis ($x = 0.5 \, \si{\micro m}$) outline of the transverse wakefields. Outlines for these same fields, obtained from effective-density-based analytical estimates (\textit{light, dashed black} lines), are provided as well. All panels show the beam density distribution (\textit{transparent-blue-green-yellow} color scale), capped (saturated) to emphasize internal perturbations caused by the CNT array layers.}
    \label{fig:w20_g20}
\end{figure*}

\begin{figure*}
    \centering
    \includegraphics[scale=1.]{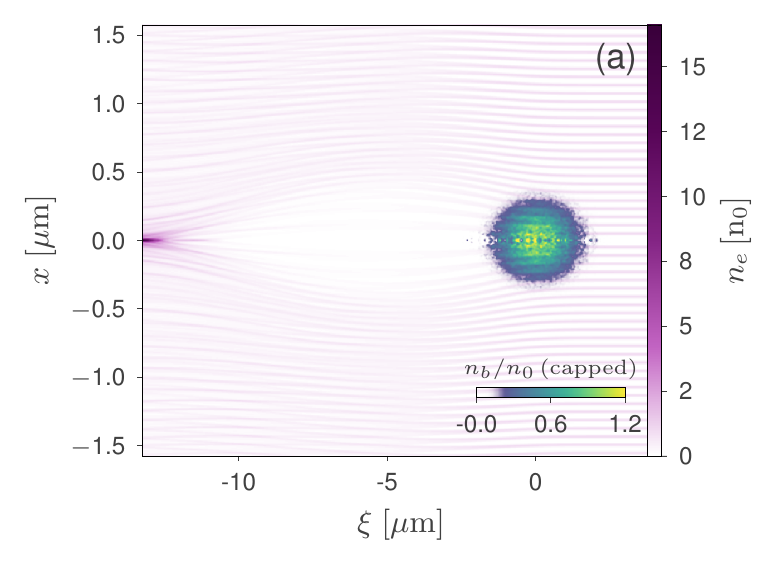}
    \includegraphics[scale=1.]{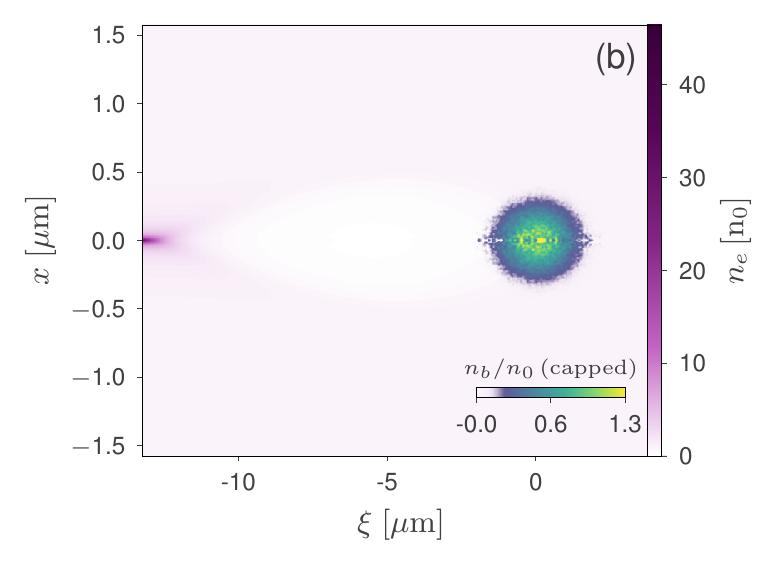} \\
    \includegraphics[scale=1.]{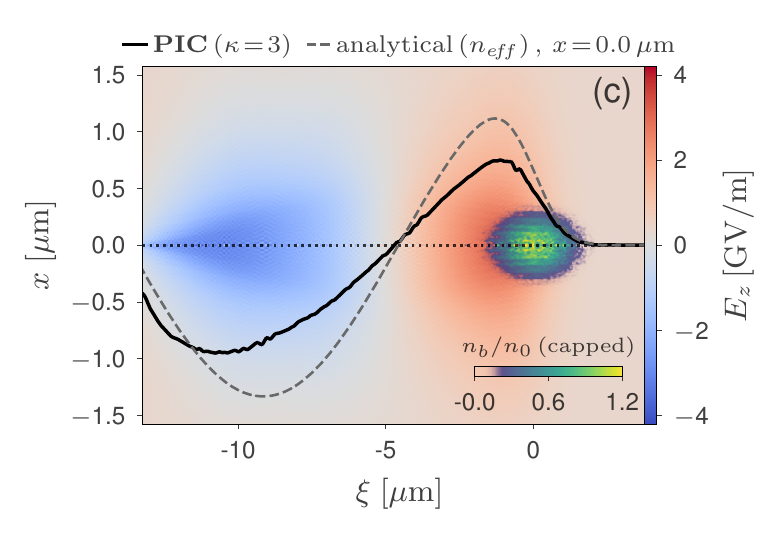}
    \includegraphics[scale=1.]{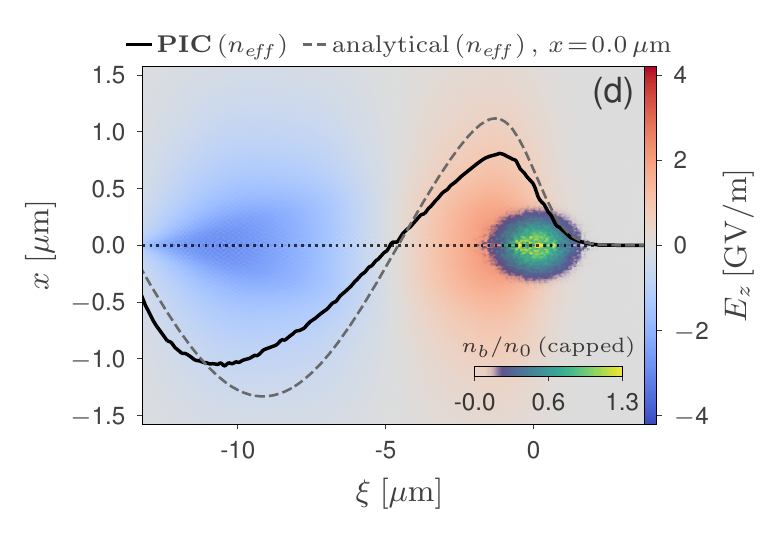} \\
    \includegraphics[scale=1.]{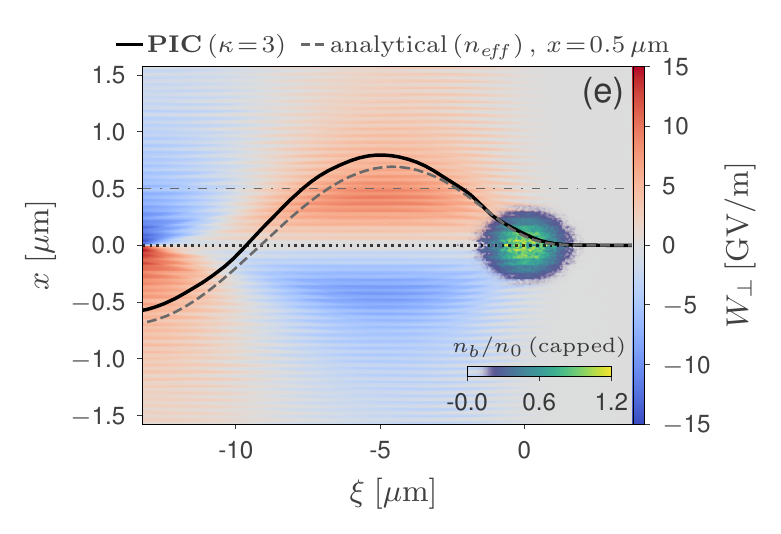}
    \includegraphics[scale=1.]{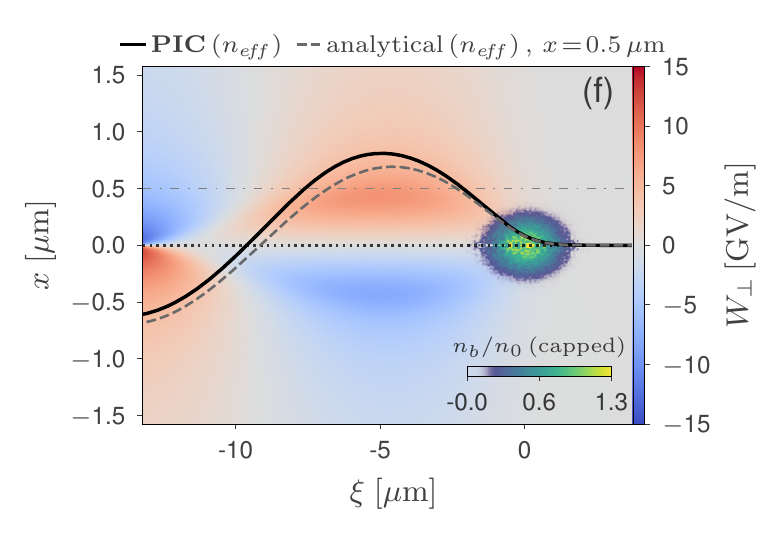}
    \caption{Simulation results for a CNT array with $w = 20 \, \si{nm}$ and $g = 40 \, \si{nm}$ ($\kappa = 3$, \textit{left column}), compared to  simulation results for a homogeneous plasma, with $n_\mathit{eff} = 0.33\,n_0$ (\textit{right column}), at a propagation distance of $z = 60 \, \si{\micro m}$. Panels (a) and (b) show the normalized plasma electron density (\textit{purple} color scale), panels (c) and (d) depict the longitudinal wakefields, and panels (e) and (f) illustrate the transverse wakefields. The fields obtained from PIC simulation results are shown as \textit{blue-grey-red} colored regions, with \textit{thick, solid black} lines representing the on-axis ($x = 0$) outline of the longitudinal wakefields, and the off-axis ($x = 0.5 \, \si{\micro m}$) outline of the transverse wakefields. Outlines for these same fields, obtained from effective-density-based analytical estimates (\textit{light, dashed black} lines), are provided as well. All panels show the beam density distribution (\textit{transparent-blue-green-yellow} color scale), capped (saturated) to emphasize internal perturbations caused by the CNT array layers.}
    \label{fig:w20_g40}
\end{figure*}

From Figs. \ref{fig:w20_g05}, \ref{fig:w20_g20}, and \ref{fig:w20_g40}, PIC simulation results for CNT arrays with a wall thickness of 20 nm and gaps of 5 nm ($\kappa = 1.25$), 20 nm ($\kappa = 2$), and 40 nm ($\kappa = 3$) can be compared against results obtained for their equivalent effective densities, $0.8\,n_0$, $0.5\,n_0$, and $0.33\,n_0$, respectively. In each of these figures, results for the CNT array are shown in the left column (panels a, c, and e), and results for its equivalent effective density are displayed in the right column (panels b, d and f). All results were taken at a propagation distance of $z = 60.5 \, \si{\micro m}$.

Figure \ref{fig:w20_g05}(a) depicts the plasma electron density (\textit{purple} color scale), in units of $n_0$, for a CNT array with wall thickness of 20 nm and gaps of 5 nm, corresponding to $\kappa = 1.25$. From this panel, one can see the fine layered-structure obtained for these parameters, with a large number of layers and gaps within the beam transverse size. Figure \ref{fig:w20_g05}(b) shows the same quantity, the plasma electron density, for a homogeneous plasma with an effective density $n_\mathit{eff} = 0.8 \, n_0$. By comparing panels (a) and (b), one can see that both cases are quantitatively and qualitatively similar. In the same figure, panels (c) and (d) show the longitudinal wakefield $E_z$, obtained for the CNT array with $\kappa = 1.25$ and for the homogeneous plasma with effective density $n_\mathit{eff} = 0.8 \, n_0$, respectively. In these panels, the \textit{blue-white-red} color scale represents $E_z (\xi,\, x,\, y=0)$, and the \textit{thick, solid black} line represents the on-axis wakefield outline,  $E_z (\xi,\, x=0,\, y=0)$. In both panels, the \textit{light, dashed black} lines represent the $E_z$ on-axis obtained from the analytical solution, given by Eq.~(\ref{eq:amplitude}), for the adopted effective density ($n_\mathit{eff} = 0.8 \, n_0$). Both $E_z$ outlines obtained from PIC simulations exhibit a small departure from the sinusoidal behaviour, observed in the analytical estimates, towards a saw-tooth-like profile. This can be explained by the fact that, for the adopted parameters, the density perturbation observed in panels (a) and (b) resembles a fully-evacuated bubble, which is a characteristic of wakefields driven in the nonlinear regime \cite{Esarey:2009}. The last two panels of Fig. \ref{fig:w20_g05} -- panels (e) and (f) -- show the transverse wakefield $W_\perp \equiv E_r - c\,B_\theta$. These panels use mostly the same line and color patterns to display quantities analog to those shown in panels (c) and (d), now related to $W_\perp$ rather than to $E_z$. The only exceptions are the transverse wakefield $W_\perp$ outlines, which are now plotted $0.5 \, \si{\micro m}$ off-axis, at the position indicated by the \textit{light, dash-dotted} lines in such panels. Since $W_\perp$ is zero on-axis, this was a necessary change. By comparing panels (c) and (e) against panels (d) and (f), respectively, one can see a remarkable agreement between the results obtained for the CNT array and the homogeneous effective density simulations, for both longitudinal and transverse wakefields.  

Figure \ref{fig:w20_g20} exhibits results for a CNT array with walls and gaps of equal size, $ w = g = 20 \, \si{\nano m}$, corresponding to $\kappa = 2$ (left column).  With this equality, achieved by setting gaps four-times wider than those of the previous case depicted in Fig. \ref{fig:w20_g05}, 50\% of the the CNT-array volume is empty. Therefore, it is remarkable that the results obtained for the equivalent homogeneous effective density $n_\mathit{eff} = 0.5 \, n_0$, displayed in the right column of Fig. \ref{fig:w20_g20}, are still in agreement with the results obtained for the CNT array, shown in the left column of the same figure. With wider gaps, the layered structure in the plasma electron density is more evident in Fig. \ref{fig:w20_g20}(a). In addition, the horizontal stripes observed in the beam density, alternating regions of higher and lower densities within the gaps and walls, respectively, show how the beam is affected by the CNT layers. Fig. \ref{fig:w20_g20}(b) shows the electron density for a homogeneous plasma with $n_\mathit{eff} = 0.5 \, \si{\micro m}$. Despite the higher peak plasma-electron-density observed in this panel ( $\sim\! 18 \, n_0$) if compared to panel (a) ($\sim\! 10 \, n_0$), the geometry of the electron density perturbation (``bubble'' radius and length), as well as the longitudinal and transverse wakefields, are quite similar in both cases. In addition, while the longitudinal wakefield depicted in Fig. \ref{fig:w20_g20}(c) is smooth and continuous, in Fig. \ref{fig:w20_g20}(e) a slightly-layered structure in the same color scale shows how the transverse wakefield is affected by the CNT-array gaps.

In Fig.~\ref{fig:w20_g40}, panels (a), (c), and (e) show results for a CNT-array with the same previously adopted wall thickness, $w = 20 \, \si{\nano m}$, and gaps twice as wide as this value, i.e., $g = 40 \, \si{\nano m}$. For this case, $\kappa = 3$, and two thirds of the CNT-array volume are empty. Simulation results for the corresponding homogeneous effective density, $n_\mathit{eff} = 0.33 \, n_0$, are displayed in panels (b), (d), and (f), in the right column of Fig.~\ref{fig:w20_g40}. Despite the gaps being predominant within the simulation domain, their more pronounced effect in the beam-density distribution and transverse wakefield, and the larger peak plasma-electron-density difference between Fig.~\ref{fig:w20_g40}(a) and Fig.~\ref{fig:w20_g40}(b), the results obtained for the equivalent homogeneous effective density are still in agreement with those obtained for the CNT array.

\begin{figure}
    \centering
    \includegraphics[scale=1.]{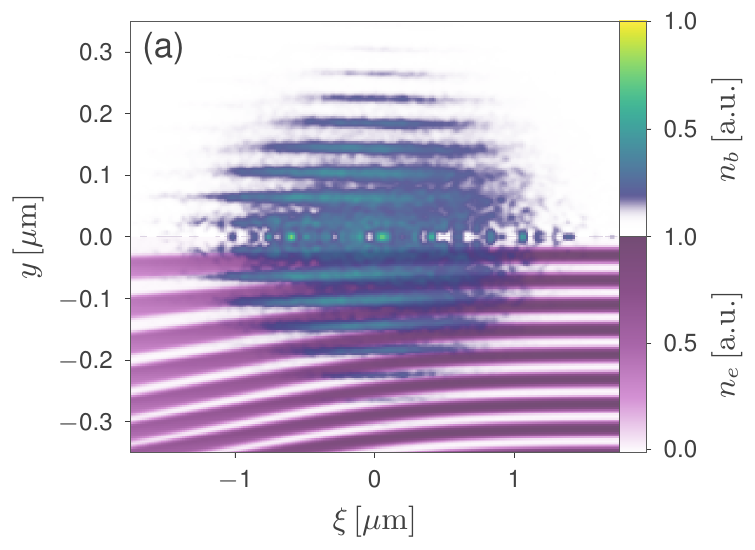}
    \includegraphics[scale=1.]{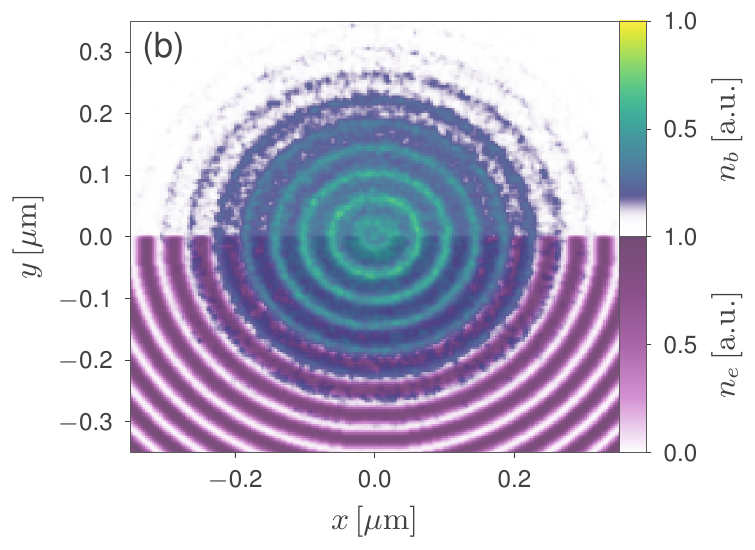}
    \caption{(a) Longitudinal and (b) transverse sections showing both the plasma electron and beam densities, $n_e$ and $n_b$, respectively, plotted in arbitrary units (a.u.) at a propagation distance of $z = 92.5 \, \si{\micro m}$. While both densities are shown in the lower-half region of each panel, only $n_b$ is displayed in their upper-half region.}
    \label{fig:cross_section}
\end{figure}

As the beam propagates in a layered structure composed by alternating high and low-density regions, the beam density is periodically modulated by such structure, being lower within the high-density regions due to the strong interference. This effect, which has already been shown by Sahai et al. \cite{Sahai:2021}, can be slightly noticed in the aforementioned beam-density modulation depicted in the CNT-array PIC simulation results (left columns of Figs. \ref{fig:w20_g05}, \ref{fig:w20_g20}, and \ref{fig:w20_g40}). In order to observe this modulation in a clearer fashion, Figs.~\ref{fig:cross_section}(a) and \ref{fig:cross_section}(b) display longitudinal and transverse slices of the simulation domain, centered at $y\!=\!0$ and $\xi\!=\!0$ respectively, containing both the beam (\textit{transparent-blue-green-yellow} color scale) and plasma-electron (\textit{purple} color scale) densities. In this figure, plotted for a CNT array with $w\!=\!g\!=\!20 \, \si{\nano m}$ ($\kappa\!=\!2$) at a propagation distance of $z\! \simeq \! 90 \, \si{\micro m}$, while the lower-half fraction of each panel shows both beam and plasma-electron densities, only the former is depicted in their upper-half region. This configuration allows for a clear view of how the beam density is low/null within the CNT array walls, and high within its gaps. The intensity of this modulation, as well as the rate at which it takes place as the beam propagates along the CNT array, will depend on the beam (initial kinetic energy, density profile) and plasma (density, gap-to-wall ratio) parameters.

\section{ \label{sec:5} Discussion and conclusions}

Depending on their particular atomic configuration and electrical conduction nature, some solid-state micro- and nanosized structures offer interesting properties to enhance electric field components or induce strong wakefields that could be useful for acceleration, as well as transverse particle guiding
and radiation emission. In particular, due to their special optoelectronic, thermal and mechanical properties, 2D materials CNT and graphene based structures might offer novel and alternative solutions to overcome present limitations of standard acceleration techniques.

In this article beam-driven CNT based solid-state plasmonic acceleration has been investigated by means of analytical calculations and numerical PIC simulations, assuming a cold relativistic fluid model, in which the motion of the plasma is not bounded or constrained. As mentioned in Sec.~\ref{sec:2}, this assumption can be a good approximation to describe the collective oscillations of a free electron gas (plasmon) in metallic nanostructures. As indicated in \cite{Ding:2020, Rider:2012} plasmons can be described by the same physics and equations of motion as plasmas. 

In CNTs, excited electrons from the conduction band can present mobility on the order of $10^4$~cm$^2/$(V$\cdot $s), practically two orders of magnitude higher than that in metals \cite{Hong:2007}. Furthermore, following previous studies \cite{Hakimi:2018}, here we have assumed that the wakefield acceleration of electrons in solid-state plasmas is practically not influenced by ionic lattice effects. Although assuming free mobility of excited plasmon electrons along a CNT based structure might be a reasonable first approximation, in future studies we need to investigate in detail the effects from the movement restrictions imposed by the solid-state properties of the ionic lattice. In the worse case scenario, these restrictions might mitigate the on-axis electron density peaks observed in the PIC simulations presented along this work. As a consequence, the amplitudes of the maximum accelerating fields, $E_z^\mathrm{max}$, which are related to these on-axis density peaks, might be mitigated as well. Therefore, the values here presented must be taken as first-order estimates for such amplitudes.

In terms of the ionic lattice, another question is the survivability of the solid-state sample interacting with the wakefield driver (a relativistic particle beam in our case). The lattice will be highly ionized by the driver and, to be effective, the accelerating wakefield must precede any lattice dissociation due to the ion motion. According to \cite{Chen:1998, Shin:2015} an ionic crystalline solid dissociates by absorbing plasmon energy on a timescale determined by $\Delta t \approx \sqrt{m_i/m_e} (2\pi/\omega_p)$, where $m_i$ and $m_e$ are the rest mass of the ion and the electron respectively. It depends on the plasma frequency $\omega_p$. Assuming plasma densities $n_0 \approx 10^{19}~\textrm{cm}^{-3}$, we obtain a lattice dissociation time $\Delta t \sim 1$~ps. As we have shown in the results of previous sections, in the wakefield excitation we are dealing with timescales $< 100$~fs. Therefore, the wakefield may build-up before the lattice dissociates.

The use of 2D carbon-based targets in the form of structured materials, such as multi-layered graphene \cite{Bontoiu:2023} and CNT bundles and arrays, as a solid-state medium for generating high-density plasmas can have important advantages over other solid materials. These structures have hollow spaces and gaps that can improve the channelling properties of the laser pulse, and also reduce scattering effects compared to bulkier materials. However, it is important to note that the interaction between high-density beams and these targets can produce a significant amount of radiation/secondary particles, as well as heating and diffusion, if collisions are taken into account. While these effects might strongly affect the beam-wall interference region, it is not clear how they would impact the electron dynamics in the gaps, which plays a relevant role in the effective density scheme. Moreover, collisions may also damp the wakefield amplitude. However, the importance of this effect will be determined by how its time scale relates to the wakefield dynamics. The evaluation of collision effects in wakefield excitation can be complex, as it involves understanding their impact on collective oscillations. Monte Carlo algorithms integrated to PIC simulation codes can be used to to assess some of these effects. However, Monte Carlo-induced numerical heating caused by (artificial) stochastic production of electromagnetic energy has been observed in such integrated simulations \cite{Alves:2021}. Hence, a comprehensive understanding on how collisions would affect the effective density scheme may require additional analysis.

For all investigated cases, single-level ($Z=1$) pre-ionized plasmas were adopted in the simulations. However, it is important to consider that variations in the preplasma, such as changes in the electron density caused by different ionization levels, may affect the properties of the beam-driven wakefield. Hence, in a more realistic scenario, the ionization process should be taken into account. For example, the ADK model \cite{Ammosov:1986,Chen:2012} could be used to ionize initially neutral targets with a laser prepulse. This could yield more realistic electron and ion distributions in the simulations, allowing for the assessment of how preplasma effects would affect the effective density scheme.
\black

In this article, the wakefield excitation in CNT based hollow solid-state structures has been studied by means of 2D axisymmetric PIC simulations using the code FBPIC \cite{Lehe:2016}. Different system configurations have been studied. First, we have investigated the case of a single tube which is basically a hollow plasma channel with a micrometric inner aperture. The channel walls are assumed to be formed of CNT bundles that will act as a solid-state plasma under the excitation and ionization driven by a external beam crossing the channel. A systematic parameter scan has been performed to study the dependence of the maximum longitudinal electric field $E^{max}_z$ on the tube aperture and wall thickness. This has allowed us to obtain analytical expressions for $E^{max}_z$ to make quick predictions. 

In the second part of this paper, a CNT array has been modeled in a 2D axisymmetric geometry. For the sake of comparison, it is worth mentioning that previously similar studies were performed considering a 2D Cartesian symmetry using the PIC code EPOCH \cite{Resta}. Due to its 2D Cartesian geometry, the system simulated in Ref.~\cite{Resta} is closer to a multilayer nanostructure alternating straight plasma layers and empty space. It could represent either multilayer graphene structures or a sequence of thick layers made of CNT bundles. Assuming similar parameters, in comparison to 2D Cartesian, the wakefield amplitudes obtained from 2D axisymmetric simulations are approximately one order of magnitude lower, matching analytical estimates from the linear theory while in the linear regime (see Fig.~\ref{fig:n_eff}).

The adoption of an effective density $n_\mathit{eff}$ allows existing analytical estimates, derived for wakefields excited in homogeneous plasmas in the linear regime, to be used for describing wakefields excited in CNT arrays. The accuracy of this approach for estimating the amplitude of the longitudinal wakefield accelerating amplitude $E_z^\mathrm{max}$ was verified as follows. PIC simulations for homogeneous plasmas with effective densities given by $n_\mathit{eff} = n_0/\kappa$ were performed for values of $\kappa$ ranging from $\kappa = 1$, representing a uniform plasma with no gaps, to $\kappa = 3$, representing CNT arrays with gaps three times wider than the CNT wall thicknesses. The accelerating field amplitudes $E_z^\mathrm{max}$ obtained from such simulations show good agreement with the aforementioned analytical estimates (Fig.~\ref{fig:n_eff_2}). Two additional sets of PIC simulations were performed, each of them assuming CNT arrays with wall thicknesses of 20 nm and 40 nm, respectively, with $\kappa$ varying from 1 to 3. From the PIC simulation results for the 20-nm CNT arrays, it can be seen that $E_z^\mathrm{max}$ shows good agreement with the analytical estimates obtained by using the effective density $n_\mathit{eff}$, all along the range of investigated $\kappa$ values. On the other hand, results for the 40-nm CNT show that this agreement is reduced as the gaps get wider with respect to the CNT wall thickness.

It is worth noting that, as the gaps (and, consequently, $\kappa$) are increased, the wavelength of the wakefield driven in the CNT array decreases accordingly with the plasma wavelength associated to the effective density $n_\mathit{eff}$. In other words, the wakefield wavelength scales with an effective plasma wavelength, $\lambda_\mathit{p,eff} = 2\pi/k_\mathit{p,eff}$. This finding may be relevant for obtaining analytical estimates for plasma-based suitable parameters for the wakefield driver, as well as limiting parameters such as the length of the accelerating/decelerating wakefield phase, or the dephasing and depletion lengths if a laser pulse is adopted as the wakefield driver.

Analytical estimates obtained by using the effective density may provide upper-bound estimates for the accelerating fields in plasmas with complex, periodic profiles. The accuracy of such estimates obtained by using this approach might depend on how large is a unitary cell of plasma periodic pattern, if compared to the beam driver size. For the largest departure from a homogeneous plasma presented in this paper, i.e., for $\kappa = 3$ ($n_\mathit{eff} = 0.33\,n_0$), the size of such a unitary cell is approximately $0.07 \, \sigma_r$ and $0.14 \, \sigma_r$ for the 20 nm and 40 nm CNT arrays, respectively. The accuracy of the effective density approach might be affected by the CNT-wall plasma density as well. While very low densities might not have enough electrons to populate the gaps, higher densities may decrease the transverse reach of the wakefield, which scales with $k_p^{-1}$. In a future work, further investigation will be conducted on the validity and limitations of the effective density approach. Future studies will also include the design of possible experiments to validate some of the aspects of results described in this article.

Eventually, nanostructured materials based on CNTs and graphene could lead to novel pathways to access multi-GV/m and multi-TV/m field regimes towards more sustainable, compact and low-cost accelerating methods. This could open new horizons to the physics of extreme fields, particularly in collider physics, light sources, and in many other areas of applied sciences, medicine and industry.

\vspace{-0.5cm}
\begin{acknowledgments}
\vspace{-0.3cm}
This project is supported by the Generalitat Valenciana (CIDEGENT/2019/058). The authors acknowledge computing resources provided by STFC’s SCARF cluster, the University of Manchester’s CSF3 cluster, the Conselho Nacional de Desenvolvimento Científico e Tecnológico do Brasil, CNPq (Chamada Universal 427273/2016-1), and the Fundação de Amparo à Pesquisa do Estado do Rio Grande Do Sul, FAPERGS (PqG 21/2551-0002027-0). Finally, the authors thank the FBPIC and OpenPMD developers for their contributions to the scientific community, and Aakash A. Sahai, for his previous collaboration with our group.
\end{acknowledgments}

\appendix


\end{document}